\documentclass[aps,reprint,onecolumn,pre,tighten,nofootinbib,eqsecnum]{revtex4-1}

\usepackage{epsfig}
\usepackage{graphicx}
\usepackage{amsmath}
\usepackage{amssymb}
\usepackage{bm}
\usepackage{mathrsfs}

\begin{document}
\title{Enskog kinetic theory of rheology for a moderately dense inertial suspension}
\author{Satoshi Takada}
\email[e-mail:]{takada@go.tuat.ac.jp}
\affiliation{Institute of Engineering, Tokyo University of Agriculture and Technology,
2--24--16, Naka-cho, Koganei, Tokyo 184--8588, Japan}
\author{Hisao Hayakawa}
\email[e-mail:]{hisao@yukawa.kyoto-u.ac.jp}
\affiliation{Yukawa Institute for Theoretical Physics, Kyoto University, Kyoto 606--8502, Japan}
\author{Andr\'es Santos}
\email[e-mail:]{andres@unex.es}
\author{Vicente Garz\'o}
\email[e-mail:]{vicenteg@unex.es}
\affiliation{Departamento de F\'{\i}sica and Instituto de Computaci\'on Cient\'{\i}fica Avanzada (ICCAEX),
Universidad de Extremadura, E--06006 Badajoz, Spain}

\begin{abstract}
The Enskog kinetic theory for moderately dense inertial suspensions under simple shear flow is considered as a model to analyze the rheological properties of the system.
The influence of the background fluid on suspended particles is modeled via a viscous drag force plus a Langevin-like term defined in terms of the background temperature.
In a previous paper [Hayakawa et al., Phys.\ Rev.\ E {\bf 96},  042903 (2017)], Grad's moment method with the aid of a linear shear-rate expansion was employed to obtain a theory which gave good agreement with the results of event-driven Langevin simulations of hard spheres for low densities and/or small shear rates.
Nevertheless, the previous approach had a limitation of applicability to the high shear-rate and high density regime.
Thus, in the present paper, we extend the  previous work and develop Grad's theory including higher order terms in the shear rate. 
This improves significantly the theoretical predictions, a  quantitative agreement between theory and simulation being found  in the high-density region (volume fractions smaller than or equal to $0.4$).
\end{abstract}
\date{\today}
\maketitle

\section{Introduction}
\label{introduction}

Shear thickening is a rheological process in which the viscosity drastically increases as the shear rate increases.
There are two types of shear thickening:
continuous shear thickening (CST) and discontinuous shear thickening (DST).
In particular, DST has been used for many industrial applications, such as protective vests and traction controls.

DST has attracted much attention among physicists~\cite{Barnes89,Mewis11,Brown14,Lootens05,Cwalina14} as a typical nonequilibrium discontinuous phase transition between a liquid-like phase and a solid-like phase.
Although most of previous studies on shear thickening are oriented to dense suspensions,
it is convenient to analyze whether kinetic theory~\cite{Brilliantov04,Brey98,Garzo99,Lutsko05,Garzo13,Garzo} can be used for a quantitative theoretical description.
Some papers have reported that a DST-like process for the kinetic temperature can take place as a result of a saddle-node bifurcation of ignited-quenched transition~\cite{Tsao95,Sangani96,BGK2016,Chamorro15,DST16,Saha17, Gonzalez20b}.
Some of the previous theories are based on a suspension model which ignores thermal fluctuations in the dynamics of grains~\cite{Tsao95,Sangani96,Saha17}. 
A more refined suspension model including a Langevin-like term has been more recently considered in Refs.\ \cite{BGK2016,Chamorro15,DST16,Hayakawa17}.
The validity of these studies can be verified by event-driven Langevin simulation for hard spheres (EDLSHS)~\cite{Scala12}.
The target systems described by the kinetic theory are inertial suspensions~\cite{Koch01}, which can be regarded as an idealistic model of aerosols.

Although the previous achievements of Refs.~\cite{Tsao95,DST16,Saha17,Sugimoto20} for low-density inertial suspensions are remarkable,
Sangani et al.~\cite{Sangani96} showed that the discontinuous transition of the kinetic temperature for dilute suspensions becomes continuous at relatively low density.
This conclusion agrees with previous theories ~\cite{BGK2016,Saha17,Hayakawa17}.
Hayakawa et al.~\cite{Hayakawa17} developed the Enskog kinetic theory associated with Grad's expansion~\cite{Grad49} to the first order in the shear rate and in the kinetic stress tensor.
Although the authors illustrated a reasonable agreement between the theoretical predictions and the results of simulations, there is a shortcoming  in the study of Ref.\ \cite{Hayakawa17}.
Since the theory is constructed within the linear approximation in the shear rate,
the previous analysis is not applicable to systems under high shear rates.
It should be noted that such effects are not important for dilute systems but they are relevent in dense systems.
Therefore, we need to extend the previous analysis to include effects of high shear rates.

The purpose of this paper is to extend the previous dilute results to moderately dense systems by solving the Enskog kinetic equation~\cite{Garzo99,Lutsko05,Garzo13,Resibois77} by two complementary and independent routes: Grad's moment method and EDLSHS.
The influence of the background fluid on grains is modeled via an external force constituted by two terms:
(i) a viscous drag force which mimics the dissipation of suspended particles with the interstitial fluid and
 (ii) a stochastic Langevin-like term accounting for thermal fluctuations.
This second term accounts for the energy transfer between grains and the background fluid.
To assess the finite-density effects on rheology, a set of coupled equations for the stress tensor, the kinetic temperature, and the anisotropic temperatures corresponding to the normal stress differences are derived from Grad's approximation.
The validity of  our simple theory is also examined through a comparison with computer simulations based on EDLSHS.
The motivation of the the present work is threefold. 
First, since there is some evidence~\cite{Garzo99,Chialvo13} that the Enskog theory is accurate for solid volume fractions smaller than $0.5$, our results  allow us to analyze the behavior of rheology for moderately dense suspensions corresponding to typical experiments.
As the second point, our results allow us to clarify whether the scenario proposed by Sangani et al.~\cite{Sangani96} is universal.
As the third point, we extend the previous work~\cite{Hayakawa17} to the theory which can describe the high shear rate and density regime.

The organization of this paper is as follows.
The outline of the Enskog kinetic theory of moderately dense suspensions under a simple shear flow and the connection between  kinetic theory and the Langevin equation are briefly summarized in Sec.\ \ref{sec:Enskogbis}, which consists of two parts.
In Sec.\ \ref{sec:Langevin}, we explain the relationship between the Enskog kinetic theory and the Langevin dynamics.
In Sec.\ \ref{sec:moment}, we summarize the moment equations which are necessary to describe the rheology.
Section \ref{sec:rheology} summarizes the theoretical results for the rheology of sheared inertial suspensions based on the Enskog kinetic theory with the aid of Grad's moment method.
That section consists of two parts.
In Sec.\ \ref{sec:Grad}, we briefly introduce Grad's moment method.
In Sec.\ \ref{sec:arbitrary_shear}, we explain the general framework to describe rheology under arbitrary shear rate and discuss the convergence of the theoretical results by checking the truncation cutoff terms.
In Sec.\ \ref{sec:simulation}, we demonstrate that the present theory gives quantitatively precise results, even for $\varphi=0.5$ in the case of the kinetic temperature, where $\varphi$ is the volume fraction of grains.
Finally,   our results are summarized and discussed in Sec.\ \ref{sec:discussion}.
Some technical parts are relegated to two Appendices.
In Appendix \ref{sec:R}, we discuss the results if the drag coefficient depends on the density.
In Appendix \ref{sec:derivation_I}, we present the detailed derivations of the collisional integrals for arbitrary shear rate.

\section{Enskog kinetic equation for suspensions under simple shear flow}\label{sec:Enskogbis}

\subsection{Langevin equation and Enskog equation}\label{sec:Langevin}

Let us consider a three-dimensional  collection of monodisperse smooth hard spheres of diameter $\sigma$, mass $m$, and restitution coefficient $e$ satisfying $0< e \le 1$.
The suspended particles are immersed in a solvent or fluid phase (fluidized inertial suspension) and are  subject to simple (or uniform) shear flow in which $x$ and $y$ are the direction of shear and the direction of velocity change, respectively (where the shear flow is symmetrical with $y=0$).
The simple shear flow is macroscopically characterized by a uniform density $n$, a uniform kinetic temperature $T$, and a macroscopic velocity field $\bm{u}=(u_x,\bm{u}_\perp)$ of the form
\begin{equation}
\label{plane_shear}
u_x=\dot\gamma y, \quad \bm{u}_\perp=\bm{0},
\end{equation}
where $\dot\gamma$ is the \emph{constant} shear rate.

As in Ref.\ \cite{Hayakawa17}, for low Reynolds numbers, the Langevin equation turns out to be a reliable model for studying the dynamic properties of the suspended particles. Neglecting the influence of gravity, the Langevin equation reads \cite{K81}
\begin{equation}
\label{Langevin_eq}
\frac{d{\bm{p}}_i}{dt}=-\zeta \bm{p}_i + \bm{F}_i^{{\rm imp}}+ m\bm{\xi}_i.
\end{equation}
Here, $\bm{p}_i\equiv m(\bm{v}_{i}-\dot\gamma y_i \bm{e}_x)$ is the peculiar momentum of $i$-th particle,  where $\bm{v}_i$ is the (instantaneous) velocity and $\bm{e}_\alpha$ is the unit vector parallel to $\alpha$-direction, $\bm{F}_i^{\rm imp}$ is the impulsive force which accounts for the grain collisions, and $\bm{\xi}_i(t)=\xi_{i,\alpha}(t)\bm{e}_\alpha$ is the noise with the statistical properties
\begin{equation}
\label{noise}
\langle \bm{\xi}_i(t)\rangle=\bm{0}, \quad
\langle \xi_{i,\alpha}(t)\xi_{j,\beta}(t')\rangle = \frac{2\zeta T_{\rm ex}}{m} \delta_{ij}\delta_{\alpha\beta}\delta(t-t').
\end{equation}
In Eq.\ \eqref{Langevin_eq}, $\zeta$ is the drag coefficient characterizing the drag from the background fluid and $T_{\rm ex}$ is the temperature of the interstitial fluid (consisting of molecular gases).

While the drag coefficient $\zeta$ should be in general a resistance matrix as a result of the hydrodynamic interactions between grains, in the case of relatively dilute suspensions it can be assumed to be a scalar  ($\zeta \propto \eta_0 \propto \sqrt{T_{\rm ex}}$, $\eta_0$ being  the viscosity of the solvent or fluid phase). 
In addition, for the sake of simplicity, throughout this paper we will regard $\zeta$ as a constant independent of density (see Appendix \ref{sec:R} for the results when we consider the dependence of the drag coefficient on the packing fraction of the grains; we find that this density dependence does not  change the results qualitatively). 
This simple model might be applicable to the description of inertial suspensions in which the mean diameter of suspended particles is approximately ranged from $1\,\mu$m to $70\,\mu$m~\cite{Koch01}.
Note that if we ignore the density dependence of $\zeta$ and the grains are bidisperse soft spheres,
the Langevin model~\eqref{Langevin_eq} is equivalent to that used by Kawasaki {et al}.~\cite{Kawasaki14}.

As said above, we assume now that the suspension is under simple shear flow.
At a microscopic level, this state is generated by Lees--Edwards boundary conditions~\cite{LE72}, which are simply periodic boundary conditions in the local Lagrangian frame moving with the flow velocity $\bm{u}$. In this reference frame, the velocity distribution function becomes uniform, i.e., $f(\bm{r},\bm{v},t)=f(\bm{V},t)$, where $\bm{V}=\bm{v}-\dot\gamma y\bm{e}_x$ is the peculiar velocity. Under these conditions, the Enskog kinetic equation for the inertial suspension becomes~\cite{Hayakawa17,Hayakawa03}
\begin{equation}
	\label{Enskog}
	\left(\frac{\partial}{\partial t}-\dot\gamma V_y\frac{\partial}{\partial V_{x}}\right)f(\bm{V},t)
	= \zeta\frac{\partial}{\partial \bm{V}} \cdot 
	\left\{\left[ \bm{V}+ \frac{T_{\rm ex}}{m} \frac{\partial}{\partial \bm{V}} \right] f(\bm{V},t) \right\}+J_\text{E}[\bm{V}|f,f].
\end{equation}
The Enskog collision operator $J_\text{E}[\bm{V}|f,f]$ is given by~\cite{Hayakawa17}
\begin{equation}
\label{J(V|f)}
J_{\text{E}}\left[\bm{V}_{1}|f,f\right] =\sigma^{2}g_0 \int d{\bf V}
_{2}\int d\widehat{\boldsymbol{\sigma}}\,\Theta (\widehat{{\boldsymbol {\sigma}}}
\cdot \bm{V}_{12})(\widehat{\boldsymbol {\sigma }}\cdot \bm{V}_{12})\left[
\frac{
f(\bm{V}_1'',t)
f(\bm{V}_2''+\dot\gamma\sigma \widehat{\sigma}_y \bm{e}_x,t)}{e^2}
-f(\bm{V}_1,t)f(\bm{V}_2-\dot\gamma\sigma \widehat{\sigma}_y \bm{e}_x,t)
\right],
\end{equation}
where $g_0$ is the radial distribution at contact for hard spheres, which is a function of the volume fraction  $\varphi=(\pi/6) n\sigma^3$. 
Here, $g_0$ is well approximated by~\cite{CS}
\begin{equation}
\label{radial_fn}
g_0(|\bm{r}|=\sigma,\varphi)=\frac{1-\varphi/2}{(1-\varphi)^3}
\end{equation}
for $\varphi< 0.49$.
In Eq.\ \eqref{J(V|f)}, $\Theta(\cdot)$ is the Heaviside step function, $\widehat{\bm{\sigma}}=(\bm{r}_2-\bm{r}_1)/\sigma$ is the unit vector pointing from particle $1$ to particle $2$,
and  $\bm{V}_{12}=\bm{V}_1-\bm{V}_2=\bm{v}_1-\bm{v}_2$ is the relative velocity at contact. Note that in Eq.\ \eqref{J(V|f)}, both $\bm{V}_1=\bm{v}_1-\dot\gamma y_1\bm{e}_x$ and $\bm{V}_2=\bm{v}_2-\dot\gamma y_1\bm{e}_x$ are referred to the flow velocity at the \emph{same} point $\bm{r}_1$, so that $\bm{v}_2-\dot\gamma y_2\bm{e}_x=\bm{V}_2-\dot\gamma\sigma \widehat{\sigma}_y \bm{e}_x$ since $y_2-y_1=\sigma \widehat{\sigma}_y$.
In addition, the double primes in Eq.\ \eqref{J(V|f)} denote the pre-collisional velocities $\left\{\bm{V}_1'', \bm{V}_2''\right\}$ that lead to $\left\{\bm{V}_1, \bm{V}_2\right\}$ following a binary collision:
\begin{equation}
\label{collision_rule}
\bm{V}_1''=\bm{V}_1-\frac{1+e}{2e}(\bm{V}_{12}\cdot\widehat{\bm{\sigma}})\widehat{\bm{\sigma}}, \quad
\bm{V}_2''=\bm{V}_2+\frac{1+e}{2e}(\bm{V}_{12}\cdot\widehat{\bm{\sigma}})\widehat{\bm{\sigma}}.
\end{equation}
In this paper we do not consider the effects of tangential friction and rotation induced by each binary collision.

Since the heat flux vector vanishes in the simple shear flow problem, the pressure tensor $\mathsf{P}$ becomes the most relevant quantity. It has kinetic and collisional transfer contributions, i.e., $\mathsf{P}=\mathsf{P}^k+\mathsf{P}^c$. The kinetic contribution is
\begin{equation}
\label{pressure_tensor:kinetic}
P^k_{\alpha\beta}=m \int d\bm{V} V_\alpha V_\beta f(\bm{V}),
\end{equation}
while its collisional contribution is given by~\cite{Santos98, Montanero99, Hayakawa17, Garzo}
\begin{equation}
	\label{pressure:collisional}
	P^c_{\alpha\beta}
	=\frac{1+e}{4} m\sigma^3 g_0 \int d\bm{V}_1 \int d\bm{V}_2\int d\widehat{\bm{\sigma}}
	\Theta(\bm{V}_{12}\cdot\widehat{\bm{\sigma}})
	(\bm{V}_{12}\cdot\widehat{\bm{\sigma}})^2\widehat{\sigma}_\alpha\widehat{\sigma}_\beta
	f\left(\bm{V}_1+\frac{1}{2}\dot\gamma \sigma \widehat{\sigma}_y \bm{e}_x\right)
	f\left(\bm{V}_2-\frac{1}{2}\dot\gamma \sigma \widehat{\sigma}_y \bm{e}_x\right).
\end{equation}
Because of the translational symmetry with respect to the velocity, the following procedures are not changed even when we choose $f(\bm{V}_1+\dot\gamma \sigma \widehat{\sigma}\bm{e}_x) f(\bm{V}_2)$ instead of $f(\bm{V}_1+\frac{1}{2}\dot\gamma \sigma \widehat{\sigma}_y \bm{e}_x) f(\bm{V}_2-\frac{1}{2}\dot\gamma \sigma \widehat{\sigma}_y \bm{e}_x)$ in Eq.\ \eqref{pressure:collisional}.
The trace of the pressure tensor defines the hydrostatic pressure as $P\equiv P_{\alpha\alpha}/3$. In what follows, we adopt Einstein's rule for the summation, {i.e.}, $P_{\alpha\alpha}=\sum_{\alpha=1}^3 P_{\alpha\alpha}$.
Note that, by definition, the kinetic part of the hydrostatic pressure satisfies the equation of state of ideal gases, namely  $P^k\equiv P^k_{\alpha\alpha}/3 =n T$, where $n=\int d\bm{V} f(\bm{V})$ is the number density and
\begin{equation}
\label{kinetic_T}
T =\frac{m}{3n}\int d\bm{V} \bm{V}^2f(\bm{V})
\end{equation}
is the kinetic temperature.

The suspension model defined by Eq.\ \eqref{Enskog} can be seen as a simplified version of the model introduced in Refs.\ \cite{Gradenigo11, Garzo12} to obtain the Navier--Stokes transport coefficients \cite{Garzo13PF, Gonzalez19}. 
In this latter model, the friction coefficient of the drag force ($\gamma_{\rm b}$ in the notation of Ref.\ \cite{Garzo13PF}) and the strength of the stochastic term ($\xi_{\rm b}^2$ in the notation of Ref.\ \cite{Garzo13PF}) are considered to be in general different. 
Here, to be consistent with the fluctuation-dissipation theorem for elastic collisions, both coefficients are related as $\xi_{\rm b}^2=2\gamma_{\rm b}T_{\rm ex}/m$. 
In addition, an extension of the suspension model to multicomponent systems \cite{Khalil13} has been recently considered to determine the Navier--Stokes transport coefficients for moderate densities \cite{Gonzalez20}.

\subsection{Moment equations}\label{sec:moment}

By multiplying both sides of Eq.\ \eqref{Enskog} by $m V_\alpha V_\beta$ and integrating over $\bm{V}$, one obtains the evolution equations for the kinetic contribution $P^k_{\alpha\beta}$ to the pressure tensor as
\begin{equation}
	\frac{\partial}{\partial t}P^k_{\alpha\beta}
	+\dot\gamma (\delta_{\alpha x}P_{y \beta}^k+\delta_{\beta x} P_{y \alpha}^k)
	=-2\zeta ( P_{\alpha\beta}^k- n T_{\rm ex} \delta_{\alpha\beta} )
	-\Lambda_{\alpha\beta},
	\label{Garzo31}
\end{equation}
where
\begin{equation}
	\label{Garzo32}
	\Lambda_{\alpha\beta} \equiv -m \int d\bm{V} V_\alpha V_\beta J_{\rm E}[\bm{V}|f,f] .
\end{equation}
The collisional moment \eqref{Garzo32} can be decomposed as
\begin{equation}
	\label{vic4}
	\Lambda_{\alpha\beta}
	=\overline{\Lambda}_{\alpha\beta}
	+\dot\gamma (\delta_{\alpha x}P_{y \beta}^c+\delta_{\beta x}
P_{y \alpha}^c),
\end{equation}
where $\overline{\Lambda}_{\alpha\beta}$ satisfies \cite{Hayakawa17}
\begin{align}
	\overline{\Lambda}_{\alpha\beta}
	&\equiv \frac{1+e}{4}m\sigma^{2}g_0 \int d\bm{V}_1 \int d\bm{V}_2 \int d\widehat{\bm{\sigma}}
	\Theta(\bm{V}_{12}\cdot\widehat{\bm{\sigma}})
	(\bm{V}_{12}\cdot\widehat{\bm{\sigma}})^2\nonumber\\
	&\hspace{1em}\times\left[V_{12,\alpha}\widehat{\sigma}_\beta + V_{12,\beta}\widehat{\sigma}_\alpha
	-(1+e)(\bm{V}_{12}\cdot\widehat{\bm{\sigma}}) \widehat{\sigma}_\alpha \widehat{\sigma}_\beta\right]
	f\left(\bm{V}_1+\dot\gamma \sigma \widehat{\sigma}_y \bm{e}_x\right)
	f\left(\bm{V}_2\right).
	\label{eq:def_over_Lambda}
\end{align}
The moment equations \eqref{Garzo31} can be rewritten in an alternative way by taking into account Eq.\ \eqref{vic4}:
\begin{equation}
	\label{Garzo31b}
	\frac{\partial}{\partial t}P^k_{\alpha\beta}
	+\dot\gamma \left(\delta_{\alpha x}P_{y \beta}+\delta_{\beta x} P_{y \alpha}\right)
	=-2\zeta \left(P_{\alpha\beta}^k- n T_{\rm ex} \delta_{\alpha\beta}\right)
	-\overline{\Lambda}_{\alpha\beta}.
\end{equation}

The simple shear flow state is in general non-Newtonian.
This can be characterized by rheological functions measuring the departure from the corresponding Navier--Stokes description.
Thus, we introduce  the differences $\Delta T$ and $\delta T$ of anisotropic temperatures  which are defined as
\begin{equation}
	\label{DT}
	\Delta T \equiv \frac{P_{xx}^k-P_{yy}^k}{n},\quad
	\delta T \equiv \frac{P_{xx}^k-P_{zz}^k}{n}.
\end{equation}
Obviously, $\delta T$ is meaningless in the two-dimensional case. 
In terms of $T$, $\Delta T$, and $\delta T$, the diagonal elements of the kinetic pressure tensor can be written as
\begin{subequations}
\label{Pxxk-Pzzk}
  \begin{eqnarray}
    P_{xx}^k&=&n\left(T+\frac{1}{3}\Delta T+\frac{1}{3}\delta T\right),
    \label{Pxxk}\\
    P_{yy}^k&=&n\left(T-\frac{2}{3}\Delta T+\frac{1}{3}\delta T\right),
    \label{Pyyk}\\
    P_{zz}^k&=&n\left(T+\frac{1}{3}\Delta T-\frac{2}{3}\delta T\right).
    \label{Pzzk}
  \end{eqnarray}
\end{subequations}
Apart from the normal stresses, one can define the apparent shear viscosity  $\eta(\dot\gamma,e)$ by
\begin{equation}
\label{shear_viscosity}
	\eta(\dot\gamma,e)\equiv -\frac{P_{xy}}{\dot\gamma}.
\end{equation}

The evolution equations for $T$, $\Delta T$, $\delta T$, and $P_{xy}^k$ can be easily derived from Eq.\ \eqref{Garzo31}.
They are given by
\begin{subequations}
\label{partT-part_P_{xy}}
\begin{eqnarray}
	\label{partT}
	\frac{\partial}{\partial t} T&=&
	-\frac{2}{d n}\dot\gamma P_{xy}^k+2\zeta (T_{\rm ex}-T) -
	\frac{{\Lambda}_{\alpha\alpha}}{3 n},
	\\
	\label{part_DT}
	\frac{\partial}{\partial t} \Delta T&=&
	-\frac{2}{n}\dot\gamma P_{xy}^k-2\zeta \Delta T-\frac{{\Lambda}_{xx}-{\Lambda}_{yy}}{n},
	\\
	\label{part_dT}
	\frac{\partial}{\partial t}\delta T&=&
	-\frac{2}{n}\dot\gamma P_{xy}^k-2\zeta \delta T-\frac{{\Lambda}_{xx}-{\Lambda}_{zz}}{n},
	\\
	\label{part_P_{xy}}
	\frac{\partial}{\partial t}P_{xy}^k&=&- \dot\gamma P_{yy}^k
	-2\zeta P_{xy}^k-{\Lambda}_{xy}.
\end{eqnarray}
\end{subequations}
The moment equations \eqref{partT-part_P_{xy}} are still exact within the framework of the Enskog equation and have been obtained without the explicit knowledge of the velocity distribution function $f(\bm{V},t)$.
By taking into account Eq.\ \eqref{Pyyk}, one has a set of equations for $T$, $\Delta T$, $\delta T$, and $P_{xy}^k$ where only those quantities appear explicitly apart from the collisional moments ${\Lambda}_{\alpha\beta}$.
Note that Eqs.\ \eqref{partT-part_P_{xy}} are equivalent to
Eqs.\ (25)--(28) of Ref.~\cite{Hayakawa17} when one performs the formal replacements  ${\Lambda}_{\alpha\beta}\to \overline{\Lambda}_{\alpha\beta}$, $\dot{\gamma}P_{xy}^k\to \dot{\gamma}P_{xy}$, and $\dot{\gamma}P_{yy}^k\to \dot{\gamma}P_{yy}$.

While formally exact, Eqs.\ \eqref{partT-part_P_{xy}} do not make a closed set due to the presence of the collisional tensor ${\mathsf{\Lambda}}$. The collisional pressure tensor $\mathsf{P}^c$ also needs to be evaluated to determine the rheological properties. Thus, an approximate closure is needed to deal with a closed set.  The difficult part is to evaluate those collisional quantities (${\mathsf{\Lambda}}$ and $\mathsf{P}^c$) under arbitrary shear rate.

To avoid such a technical difficulty, Grad's approximation was adopted in Ref.\ \cite{Hayakawa17}, although only linear terms in the shear rate were accounted for in those calculations.
On the other hand, some previous papers \cite{Santos98, Montanero99} obtained the complete expression of  $\mathsf{P}^c$ in Grad's approximation under arbitrary shear rate. In the present paper, we revisit the study carried out in Ref.\ \cite{Hayakawa17} and explicitly determine  ${\mathsf{\Lambda}}$ and $\mathsf{P}^c$ for arbitrary values of the shear rate $\dot{\gamma}$.

For further calculation, let us introduce $I^{(\ell)}\left(\widehat{\bm{\sigma}}\right)$ and $I_\alpha^{(\ell)}\left(\widehat{\bm{\sigma}}\right)$ as
\begin{subequations}
\label{eq:I_alpha-eq:I}
\begin{align}
	I^{(\ell)}\left(\widehat{\bm{\sigma}}\right)
	&\equiv \int d\bm{V}_1 \int d\bm{V}_2
	\Theta\left(\widehat{\bm{\sigma}}\cdot \bm{V}_{12}\right) \left(\widehat{\bm{\sigma}}\cdot \bm{V}_{12}\right)^\ell
	f(\bm{V}_1 + \dot\gamma \sigma \widehat{\sigma}_y \bm{e}_x) f(\bm{V}_2)\nonumber \\
    &= \int d\bm{V}_1 \int d\bm{V}_2
	\Theta\left(\widehat{\bm{\sigma}}\cdot \bm{V}_{12}-b\right) \left(\widehat{\bm{\sigma}}\cdot \bm{V}_{12}-b\right)^\ell
	f(\bm{V}_1 ) f(\bm{V}_2),
	\label{eq:I}\\
	\label{eq:I_alpha}
	I_\alpha^{(\ell)}\left(\widehat{\bm{\sigma}}\right)
	&\equiv \int d\bm{V}_1 \int d\bm{V}_2
	\Theta\left(\widehat{\bm{\sigma}}\cdot \bm{V}_{12}\right) \left(\widehat{\bm{\sigma}}\cdot \bm{V}_{12}\right)^\ell V_{12,\alpha}
	f(\bm{V}_1 + \dot\gamma \sigma \widehat{\sigma}_y \bm{e}_x) f(\bm{V}_2)\nonumber \\
    &= \int d\bm{V}_1 \int d\bm{V}_2
	\Theta\left(\widehat{\bm{\sigma}}\cdot \bm{V}_{12}-b\right) \left(\widehat{\bm{\sigma}}\cdot \bm{V}_{12}-b\right)^\ell \left(V_{12,\alpha}-a\delta_{\alpha x}\right)
	f(\bm{V}_1 ) f(\bm{V}_2).	
\end{align}
\end{subequations}
In the second equalities, we have made the change of variable $\bm{V}_1\to \bm{V}_1 + \dot\gamma \sigma \widehat{\sigma}_y \bm{e}_x$ and have introduced the short-hand notation
\begin{equation}
	a\equiv \dot\gamma \sigma \widehat{\sigma}_y,\quad
	b\equiv a \widehat{\sigma}_x =\dot\gamma \sigma \widehat{\sigma}_x \widehat{\sigma}_y.
\end{equation}
It should be noted that the relation $\widehat{\sigma}_\alpha I_\alpha^{(\ell)}(\widehat{\bm{\sigma}})=I^{(\ell+1)}(\widehat{\bm{\sigma}})$ is satisfied.
Using these quantities, the explicit expressions of $P_{\alpha\beta}^c$ in Eq.\ \eqref{pressure:collisional} and $\overline{\Lambda}_{\alpha\beta}$ in Eq.\ \eqref{eq:def_over_Lambda} can be rewritten as
\begin{subequations}
\label{eq:Pc-eq:overline_Lambda^E}
 \begin{eqnarray}
	P_{\alpha\beta}^c
	&=& \frac{1+e}{4}m\sigma^3 g_0 \int d\widehat{\bm{\sigma}} \widehat{\sigma}_\alpha \widehat{\sigma}_\beta I^{(2)}\left(\widehat{\bm{\sigma}}\right),
	\label{eq:Pc}\\
	\label{eq:overline_Lambda^E}
	\overline{\Lambda}_{\alpha\beta}
	&=& \frac{1+e}{4} m \sigma^{2} g_0 \int d\widehat{\bm{\sigma}}
	\left[\widehat{\sigma}_\alpha I_\beta^{(2)}\left(\widehat{\bm{\sigma}}\right) + \widehat{\sigma}_\beta I_\alpha^{(2)}\left(\widehat{\bm{\sigma}}\right)
	- (1+e)\widehat{\sigma}_\alpha \widehat{\sigma}_\beta I^{(3)}\left(\widehat{\bm{\sigma}}\right)\right].
\end{eqnarray}
\end{subequations}
Let us also introduce the tensors
\begin{subequations}
\label{Lab-Mab}
\begin{align}
\label{Lab}
	L_{\alpha \beta} &\equiv
	\int d\widehat{\bm{\sigma}}
	\left[
	\widehat{\sigma}_\alpha I_\beta^{(2)}\left(\widehat{\bm{\sigma}}\right) + \widehat{\sigma}_\beta I_\alpha^{(2)}\left(\widehat{\bm{\sigma}}\right)
		-2\widehat{\sigma}_\alpha \widehat{\sigma}_\beta I^{(3)}\left(\widehat{\bm{\sigma}}\right)
+a\left(\delta_{\alpha x} \widehat{\sigma}_\beta + \delta_{\beta x} \widehat{\sigma}_\alpha\right)I^{(2)}\left(\widehat{\bm{\sigma}}\right)
	\right],\\
	\label{Mab}
M_{\alpha \beta} &\equiv
	\int d\widehat{\bm{\sigma}} \widehat{\sigma}_\alpha \widehat{\sigma}_\beta I^{(3)}\left(\widehat{\bm{\sigma}}\right).
\end{align}
\end{subequations}
In terms of them, the expression of $\Lambda_{\alpha\beta}$ in Eq.\ \eqref{Garzo32} can be rewritten as
\begin{align}
	\Lambda_{\alpha\beta}
	= \frac{1+e}{4} m \sigma^{2} g_0 \left[L_{\alpha \beta} + (1-e)M_{\alpha \beta}\right].
	\label{eq:Lambda_E2}
\end{align}

In order to determine the tensors $\mathsf{\Lambda}$ and $\mathsf{P}^c$, only the quantities  $I^{(2)}$, $I^{(3)}$, and $I_\alpha^{(2)}$ need to be evaluated.
Apart from the lack of knowledge of the velocity distribution function $f(\bm{V})$, an extra difficulty in the evaluation of $I_\alpha^{(\ell)}$ lies in the fact that two vector geometries compete in Eq.\ \eqref{eq:I_alpha}: that of the   shearing Cartesian representation $\{\bm{e}_x,\bm{e}_y,\bm{e}_z\}$ and that of the unit vector $\widehat{\bm{\sigma}}$.
To overcome the latter difficulty, we introduce an alternative orthonormal basis $\{\bar{\bm{e}}_1,\bar{\bm{e}}_2,\bar{\bm{e}}_3\}$ defined as $\bar{\bm{e}}_i=U_{\alpha i}\bm{e}_\alpha$, where  the change of basis matrix is
\begin{equation}
\label{Uai}
	\begin{pmatrix}
		U_{x1} &U_{x2}&U_{x3}\\
		U_{y1} &U_{y2}&U_{y3}\\
		U_{z1} &U_{z2}&U_{z3}
	\end{pmatrix}
	=
	\begin{pmatrix}
		\displaystyle \frac{\widehat{\sigma}_y}{\sqrt{\widehat{\sigma}_x^2+\widehat{\sigma}_y^2}} &
		\displaystyle \frac{\widehat{\sigma}_x\widehat{\sigma}_z}{\sqrt{\widehat{\sigma}_x^2+\widehat{\sigma}_y^2}} &
		\widehat{\sigma}_x\\
		\displaystyle -\frac{\widehat{\sigma}_x}{\sqrt{\widehat{\sigma}_x^2+\widehat{\sigma}_y^2}} &
		\displaystyle \frac{\widehat{\sigma}_y\widehat{\sigma}_z}{\sqrt{\widehat{\sigma}_x^2+\widehat{\sigma}_y^2}} &
		\widehat{\sigma}_y\\
		0 &
		-\sqrt{\widehat{\sigma}_x^2+\widehat{\sigma}_y^2} &
		\widehat{\sigma}_z
	\end{pmatrix},
\end{equation}
so that $\bm{V}_{12}=V_{12,\alpha}\bm{e}_\alpha=\overline{V}_{12,i}\bar{\bm{e}}_i$ with $V_{12,\alpha}=U_{\alpha i}\overline{V}_{12,i}$ and $\overline{V}_{12,i}=U_{\alpha i}V_{12,\alpha}$. 
Note that $\bar{\bm{e}}_3=\widehat{\bm{\sigma}}$ and thus $\widehat{\bm{\sigma}}\cdot \bm{V}_{12}= \overline{V}_{12,3}$. 
We note that the Greek and Latin characters represent $\{x,y,z\}$ and $\{1,2,3\}$, respectively.

Now we define the quantities
\begin{subequations}
\label{Ja}
\begin{align}
	\bar{I}_i^{(\ell)}\left(\widehat{\bm{\sigma}}\right)
	&\equiv \int d\bm{V}_1 \int d\bm{V}_2
	\Theta\left(\overline{V}_{12,3}-b\right) \left(\overline{V}_{12,3}-b\right)^\ell \overline{V}_{12,i}
	f(\bm{V}_1) f(\bm{V}_2),\quad (i=1,2),\\
	J_\alpha\left(\widehat{\bm{\sigma}}\right)&
	\equiv U_{\alpha 1}\bar{I}_1^{(2)}\left(\widehat{\bm{\sigma}}\right)+U_{\alpha 2}\bar{I}_2^{(2)}\left(\widehat{\bm{\sigma}}\right).
\end{align}
\end{subequations}
Next, according to the definition \eqref{eq:I_alpha}, the vector $I_\alpha^{(2)}$ can be expressed in terms of $J_\alpha$, $I^{(2)}$, and $I^{(3)}$ as
\begin{equation}
	I_\alpha^{(2)}\left(\widehat{\bm{\sigma}}\right)
	=J_\alpha\left(\widehat{\bm{\sigma}}\right)  + \widehat{\sigma}_\alpha I^{(3)}\left(\widehat{\bm{\sigma}}\right)+a\left(\widehat{\sigma}_\alpha\widehat{\sigma}_x- \delta_{\alpha x}\right)I^{(2)}\left(\widehat{\bm{\sigma}}\right).
	\label{eq:relation_I2_I3_J}
\end{equation}
Since $\widehat{\sigma}_\alpha I_\alpha^{(2)}=I^{(3)}$, one has $\widehat{\sigma}_\alpha J_\alpha=0$. Inserting Eq.\ \eqref{eq:relation_I2_I3_J} into Eq.\ \eqref{Lab}, we obtain the result
\begin{equation}
\label{Lab_2}
	L_{\alpha \beta}=
	\int d\widehat{\bm{\sigma}}
	\left[
	\widehat{\sigma}_\alpha J_\beta\left(\widehat{\bm{\sigma}}\right) + \widehat{\sigma}_\beta J_\alpha\left(\widehat{\bm{\sigma}}\right)
	+2b \widehat{\sigma}_\alpha \widehat{\sigma}_\beta I^{(2)}\left(\widehat{\bm{\sigma}}\right)\right].
\end{equation}

\section{Rheology of sheared inertial suspensions via Grad's moment method}\label{sec:rheology}

\subsection{Grad's moment method}\label{sec:Grad}

In Sec.\ \ref{sec:Enskogbis}, we have presented the formal exact relations within Enskog's approximation. On the other hand, the moment equations \eqref{partT-part_P_{xy}} cannot be solved without explicit expressions for the collisional integrals ${\Lambda}_{\alpha\beta}$,
and the same applies to the collisional transfer contribution to the pressure tensor
$P_{\alpha\beta}^c$ [see Eqs.~\eqref{eq:I_alpha-eq:I}--\eqref{eq:Lambda_E2}].
Good estimates of those collisional quantities can be expected by using Grad's thirteen-moment approximation~\cite{Garzo13,BGK2016,Hayakawa17,Chamorro15,Grad49,Garzo02,Santos04}
\begin{equation}
	\label{Grad}
	f(\bm{V})=f_{\rm M}(\bm{V})\left(1+\frac{m}{2T}\Pi_{\alpha\beta}V_\alpha V_\beta\right),
\end{equation}
where
\begin{equation}
	\label{Maxwell}
	f_{\rm M}(\bm{V})=
	n\left(\frac{m}{2\pi T}\right)^{3/2} \exp\left(-\frac{m V^2}{2T} \right)
\end{equation}
is the Maxwellian distribution and
\begin{equation}
	\label{vic5}
	\Pi_{\alpha\beta} \equiv \frac{P^k_{\alpha\beta}}{nT}-\delta_{\alpha\beta}
\end{equation}
is the traceless part of the (dimensionless) kinetic pressure tensor $P^k_{\alpha\beta}$.
In principle, more terms (both isotropic and anisotropic) might be included in Eq.\ (3.1) as an expansion in Sonine or Hermite polynomials, in the spirit of Grad's method.
 In particular, this would allow us to account for the excess kurtosis associated with inelasticity ($e<1$) \cite{Hayakawa03}.
Nevertheless, it is not necessary to consider those corrections, because, as will be shown in Sec. \ref{sec:simulation}, we will get very precise results without them, at least, for $e\lesssim 1$.  

\subsection{The framework for arbitrary shear rate}\label{sec:arbitrary_shear}

The  analysis performed within the linear shear rate approximation in Ref.~\cite{Hayakawa17} is simple but in principle it only applies to low shear rates.
In this section,  we extend our previous calculations by addressing situations where the magnitude of the shear rate is arbitrary.

First, let us rewrite Eqs.\ \eqref{partT-part_P_{xy}} in dimensionless form by introducing $\theta\equiv T/T_{\rm ex}$, $\Delta \theta\equiv \Delta T/T_{\rm ex}$, and $\delta \theta\equiv \delta T/T_{\rm ex}$, 
apart from the dimensionless shear rate $\dot\gamma^*\equiv \dot\gamma/\zeta$, the scaled  time  $\tau \equiv t \zeta$, the scaled collisional moment $\Lambda_{\alpha\beta}^*\equiv \Lambda_{\alpha\beta}/(n\zeta T_{\rm ex})$, and the scaled kinetic stress tensor $\Pi_{\alpha\beta}^*=\theta \Pi_{\alpha\beta}=P_{\alpha\beta}^k/(nT_{\rm ex})-\theta\delta_{\alpha\beta}$. As a consequence, Eqs.\ \eqref{partT-part_P_{xy}} become
\begin{subequations}
\label{eq:evol_theta-eq:evol_Pi_xy}
\begin{align}
	\partial_\tau \theta
	&=  - \frac{2}{3}\dot\gamma^*  \Pi_{xy}^*+2(1-\theta)-\frac{1}{3}\Lambda_{\alpha\alpha}^*,\label{eq:evol_theta}\\
	\partial_\tau \Delta \theta
	&=  -2\dot\gamma^* \Pi_{xy}^* -2\Delta\theta-\delta \Lambda^*_{xx}+\delta\Lambda_{yy}^*,\\
	\partial_\tau \delta\theta
	&= -2\dot\gamma^*  \Pi_{xy}^* -2\delta\theta
	-2\delta \Lambda^*_{xx} - \delta\Lambda_{yy}^*,\\
	\partial_\tau \Pi_{xy}^*
	&= \dot\gamma^*  \left(\theta+\Pi_{yy}^*\right)-2\Pi_{xy}^*- \Lambda_{xy}^*,\label{eq:evol_Pi_xy}
\end{align}
\end{subequations}
with
\begin{equation}\label{delta_Lambda}
	\delta \Lambda_{xx}^*=\Lambda_{xx}^*-\frac{1}{3}\Lambda_{\alpha\alpha}^*,\quad
	\delta \Lambda_{yy}^*=\Lambda_{yy}^*-\frac{1}{3}\Lambda_{\alpha\alpha}^*.
\end{equation}
On account of Eq.\ \eqref{Pyyk},    one must insert the identity $\Pi_{yy}^*=\frac{1}{3}\delta \theta-\frac{2}{3}\Delta \theta$ in Eq.\ \eqref{eq:evol_Pi_xy}.
Note also that $\Pi_{xx}^*$ satisfies the relation $\Pi_{xx}^*=\frac{1}{3}\Delta \theta+\frac{1}{3}\delta \theta$.

From Eqs.\ \eqref{eq:Pc}, \eqref{eq:Lambda_E2}, \eqref{Mab}, and \eqref{Lab_2} we have the relations
\begin{subequations}
	\label{eq:Pc*-Lambda*}
\begin{align}
	P^{c*}
	&\equiv \frac{P_{\alpha\alpha}^c}{3nT_{\text{ex}}}=\frac{1+e}{\pi}\varphi g_0\theta 
	\int d\widehat{\bm{\sigma}} \tilde{I}^{(2)}\left(\widehat{\bm{\sigma}}\right),\\
	\Pi_{\alpha\beta}^{c*}
	&\equiv\frac{P_{\alpha\beta}^{c}}{nT_{\text{ex}}}-P^{c*}\delta_{\alpha\beta}=\frac{3}{\pi}(1+e)\varphi g_0\theta
	\int d\widehat{\bm{\sigma}} \left(\widehat{\sigma}_\alpha \widehat{\sigma}_\beta
	-\frac{1}{3}\delta_{\alpha\beta}\right)\tilde{I}^{(2)}\left(\widehat{\bm{\sigma}}\right),\label{eq:Pi_c*}\\
	\Lambda_{\alpha\beta}^*
	&= \frac{3\sqrt{2}}{\pi}(1+e)\varphi g_0 \xi_{\rm ex}\theta^{3/2} 
	\left[\widetilde{L}_{\alpha \beta} + (1-e)\widetilde{M}_{\alpha \beta}\right],
	\label{eq:Lambda*}
\end{align}
\end{subequations}
where
\begin{subequations}
\label{Lab*-Mab*}
\begin{align}
	\label{Lab*}
	\widetilde{L}_{\alpha \beta} &\equiv \frac{L_{\alpha\beta}}{nv_T^3}
	=\int d\widehat{\bm{\sigma}}
	\left[
	\widehat{\sigma}_\alpha \tilde{J}_\beta\left(\widehat{\bm{\sigma}}\right) 
	+ \widehat{\sigma}_\beta \tilde{J}_\alpha\left(\widehat{\bm{\sigma}}\right)
	+2b_T \widehat{\sigma}_\alpha \widehat{\sigma}_\beta \tilde{I}^{(2)}\left(\widehat{\bm{\sigma}}\right)\right],
	\\
	\label{Mab*}
	\widetilde{M}_{\alpha \beta} &\equiv \frac{M_{\alpha\beta}}{nv_T^3}
	= \int d\widehat{\bm{\sigma}} \widehat{\sigma}_\alpha \widehat{\sigma}_\beta 
	\tilde{I}^{(3)}\left(\widehat{\bm{\sigma}}\right).
\end{align}
\end{subequations}
Here, $\tilde{J}_\alpha\equiv J_\alpha/n^2 v_T^3$, $\tilde{I}^{(\ell)}\equiv I^{(\ell)}/n^2v_T^\ell$, and
\begin{equation}\label{3.23}
	b_T \equiv \frac{b}{v_T}=\dot\gamma \sigma  \sqrt{\frac{m}{2T}}\widehat{\sigma}_x \widehat{\sigma}_y
	= \frac{\dot\gamma^*}{\xi_{\rm ex}\sqrt{2\theta}}\widehat{\sigma}_x \widehat{\sigma}_y ,
\end{equation}
where $v_T\equiv \sqrt{2T/m}$.
The final expression of Eq.~\eqref{3.23} gives the definition of $\xi_{\rm ex}\equiv \sqrt{T_{\rm ex}/m}(\sigma \zeta)^{-1}$.
Note that the traces of $\widetilde{\sf{L}}$ and $\widetilde{\sf{M}}$ are expressed as
 $\widetilde{L}_{\alpha \alpha} =2\int d\widehat{\bm{\sigma}}b_T \tilde{I}^{(2)}\left(\widehat{\bm{\sigma}}\right)$ and $\widetilde{M}_{\alpha \alpha} =\int d\widehat{\bm{\sigma}}\tilde{I}^{(3)}\left(\widehat{\bm{\sigma}}\right)$.

Inserting Grad's approximation \eqref{Grad} into the definitions \eqref{eq:I} and \eqref{Ja}, one finally obtains\footnote{Equation \eqref{eq:I2} was given in Ref.\ \cite{Santos98} without a detailed derivation. Equation \eqref{eq:I3} was given in Ref.\ \cite{Montanero99}, but in that reference the correction $\text{erf}\to-\text{erfc}$ needs to be made. Equations \eqref{eq:Jx}--\eqref{eq:Jz} are new.} (see the detailed derivation in Appendix \ref{sec:derivation_I})
\begin{subequations}
\label{I2-I3-Jx-Jy}
\begin{align}
	\label{eq:I2}
	\tilde{I}^{(2)}(\widehat{\bm{\sigma}})
	=& -\frac{b_T}{\sqrt{2\pi}}e^{-b_T^2/2} + \frac{1+b_T^2}{2} {\rm erfc}\left(\frac{b_T}{\sqrt{2}}\right)
	+\frac{1}{2} {\rm erfc}\left(\frac{b_T}{\sqrt{2}}\right) \widehat{\sigma}_\beta \widehat{\sigma}_\gamma \Pi_{\beta \gamma}
	+\frac{b_T}{8\sqrt{2\pi}}e^{-b_T^2/2}\left(\widehat{\sigma}_\beta \widehat{\sigma}_\gamma \Pi_{\beta \gamma}\right)^2,\\
	\label{eq:I3}
	\tilde{I}^{(3)}(\widehat{\bm{\sigma}})
	=& \frac{2+b_T^2}{\sqrt{2\pi}}e^{-b_T^2/2} -\frac{1}{2}b_T \left(3+b_T^2\right) {\rm erfc}\left(\frac{b_T}{\sqrt{2}}\right)
	+3\left[\frac{e^{-b_T^2/2}}{\sqrt{2\pi}} - \frac{b_T}{2}{\rm erfc}\left(\frac{b_T}{\sqrt{2}}\right)\right]
	\widehat{\sigma}_\beta \widehat{\sigma}_\gamma \Pi_{\beta \gamma}\nonumber\\
	&+\frac{3}{8\sqrt{2\pi}}e^{-b_T^2/2}\left(\widehat{\sigma}_\beta \widehat{\sigma}_\gamma \Pi_{\beta \gamma}\right)^2,\\
	\label{eq:Jx}
	\tilde{J}_x (\widehat{\bm{\sigma}})
	=& -\widehat{\sigma}_x
	\left(\widehat{\sigma}_\beta \widehat{\sigma}_\gamma \Pi_{\beta\gamma} - \Pi_{xx} - \frac{\widehat{\sigma}_y}{\widehat{\sigma}_x}\Pi_{xy}\right)
	\left[\sqrt{\frac{2}{\pi}} e^{-b_T^2/2}\left(1+\frac{1}{4}\widehat{\sigma}_\beta \widehat{\sigma}_\gamma \Pi_{\beta\gamma}\right) - b_T {\rm erfc}\left(\frac{b_T}{\sqrt{2}}\right)
	 \right],\\
	\label{eq:Jy}
	\tilde{J}_y (\widehat{\bm{\sigma}})
	=& -\widehat{\sigma}_y
	\left(\widehat{\sigma}_\beta \widehat{\sigma}_\gamma \Pi_{\beta\gamma} - \Pi_{yy} - \frac{\widehat{\sigma}_x}{\widehat{\sigma}_y}\Pi_{xy}\right)
	\left[\sqrt{\frac{2}{\pi}} e^{-b_T^2/2}\left(1+\frac{1}{4}\widehat{\sigma}_\beta \widehat{\sigma}_\gamma \Pi_{\beta\gamma}\right) - b_T {\rm erfc}\left(\frac{b_T}{\sqrt{2}}\right)
	 \right],\\
	\label{eq:Jz}
	\tilde{J}_z (\widehat{\bm{\sigma}})
	=& -\widehat{\sigma}_z
	\left(\widehat{\sigma}_\beta \widehat{\sigma}_\gamma \Pi_{\beta\gamma} - \Pi_{zz} \right)
	\left[\sqrt{\frac{2}{\pi}} e^{-b_T^2/2}\left(1+\frac{1}{4}\widehat{\sigma}_\beta \widehat{\sigma}_\gamma \Pi_{\beta\gamma}\right) - b_T {\rm erfc}\left(\frac{b_T}{\sqrt{2}}\right)
	 \right],
\end{align}
\end{subequations}
where ${\rm erfc}(x)\equiv (2/\sqrt{\pi})\int_x^\infty dz e^{-z^2}$ is the complementary error function.
Note that, since $\text{erfc}(x)=1-\text{erf}(x)$ and $\text{erf}(-x)=-\text{erf}(x)$,  one has that $\tilde{I}^{(2)}$ is an odd function of the shear rate except for a constant and a quadratic term, $\tilde{I}^{(3)}$ is an even function except for a linear and a cubic term, and $\tilde{J}_\alpha$ is an even function except for a linear term.
 As a consequence, and taking into account Eqs.\ \eqref{delta_Lambda} and \eqref{eq:Pc*-Lambda*}, it turns out that the matrix $\mathsf{\Pi}^{c*}$ is an odd function of the shear rate except for a constant and a quadratic term (the latter only for the diagonal elements), while the matrix $\mathsf{\Lambda}^*$ is an even function except for a linear and a cubic term (the latter only for the off-diagonal element).

For given values of $e$, $\varphi$, $\dot{\gamma}^*$, and $\xi_{\text{ex}}$, the angle integrals in Eqs.\ \eqref{Lab*-Mab*} can be evaluated numerically as functions of $\theta$ (note that $b_T\propto 1/\sqrt{\theta}$), $\Delta \theta$, $\delta \theta$, and $\Pi_{xy}$. This allows us to numerically obtain the time evolution, as well as the steady-state values, of the stress tensor from the set of Eqs.\ \eqref{eq:evol_theta-eq:evol_Pi_xy} with the aid of Eqs.\ \eqref{Lab*-Mab*} and \eqref{I2-I3-Jx-Jy}.
Note that Eq.\ \eqref{eq:evol_theta-eq:evol_Pi_xy} is a closed set of equations for the kinetic stress, while the collisional contribution of the stress is given by Eq.\ \eqref{eq:Pi_c*} separately. 
Therefore, the procedure to obtain the stress is to solve the steady version of Eqs.\ \eqref{eq:evol_theta-eq:evol_Pi_xy} at first, and then evaluate Eq.\ \eqref{eq:Pi_c*} later. 
Then, we obtain the apparent viscosity from Eq.\ \eqref{shear_viscosity} with the aid of the total stress $P_{xy}$.
On the other hand, since this scheme is seen to consume too much computation time,  we employ here an alternative perturbation scheme.

\begin{table}[htbp]
	\caption{Coefficients appearing in Eqs.\ \eqref{eq:Lii_exp}--\eqref{eq:Lyy_exp} up to order $N_{\text{c}}=6$. 	\label{fig:coeff}}
\begin{ruledtabular}
 	\begin{tabular}{l}
 	$\displaystyle \tilde{\Lambda}_{\alpha\alpha}^{(0)*}
	=\frac{24}{\sqrt{\pi}}(1-e^2)\left[1+\frac{1}{20\theta^2}\left(\Pi_{xy}^{*2}+\Pi_{xx}^{*2}+\Pi_{xx}^*\Pi_{yy}^*+\Pi_{yy}^{*2}\right)\right]$\\
	$\displaystyle \tilde{\Lambda}_{\alpha\alpha}^{(1)*}
	=-\frac{4}{5\theta}(1+e)(1-3e)\Pi_{xy}^*$\\
	$\displaystyle \tilde{\Lambda}_{\alpha\alpha}^{(2)*}
	 =-\frac{2}{5\sqrt{\pi}}(1+e)(1+3e)\left[1+\frac{1}{7\theta}(\Pi_{xx}^*+\Pi_{yy}^*)-\frac{1}{28\theta^2}\Pi_{xy}^{*2}
		-\frac{1}{84\theta^2}\left(\Pi_{xx}^{*2}+\Pi_{xx}^*\Pi_{yy}^*+\Pi_{yy}^{*2}\right)\right]$\\
	$\displaystyle \tilde{\Lambda}_{\alpha\alpha}^{(3)*}=0$\\
	$\displaystyle \tilde{\Lambda}_{\alpha\alpha}^{(4)*}
	 =-\frac{1}{420\sqrt{\pi}}(1+e)(5+3e)\left[1-\frac{2}{11\theta}(\Pi_{xx}^*+\Pi_{yy}^*)+\frac{75}{572\theta^2}\Pi_{xy}^{*2}
		+\frac{3}{572\theta^2}\left(7\Pi_{xx}^{*2}+9\Pi_{xx}^*\Pi_{yy}^*+7\Pi_{yy}^{*2}\right)\right]$\\
	$\displaystyle \tilde{\Lambda}_{\alpha\alpha}^{(5)*}=0$\\
	$\displaystyle \tilde{\Lambda}_{\alpha\alpha}^{(6)*}
	 =\frac{1}{48048\sqrt{\pi}}(1+e)(3+e)\left[1-\frac{3}{5\theta}(\Pi_{xx}^*+\Pi_{yy}^*)+\frac{49}{68\theta^2}\Pi_{xy}^{*2}
		+\frac{1}{68\theta^2}\left(13\Pi_{xx}^{*2}+19\Pi_{xx}^*\Pi_{yy}^*+13\Pi_{yy}^{*2}\right)\right]$\\
	\hline
	$\displaystyle \tilde{\Lambda}_{xy}^{(0)*}
	=\frac{24}{5\sqrt{\pi}\theta}(1+e)(3-e)\Pi_{xy}^*\left[1+\frac{1}{14\theta}\left(\Pi_{xx}^*+\Pi_{yy}^*\right)\right]$\\
	$\displaystyle \tilde{\Lambda}_{xy}^{(1)*}
	=-\frac{2}{5}(1+e)\left[1-3e+\frac{2}{7\theta}(4-3e)\left(\Pi_{xx}^*+\Pi_{yy}^*\right)\right]$\\
	$\displaystyle \tilde{\Lambda}_{xy}^{(2)*}
	=\frac{2}{35\sqrt{\pi}\theta}(1+e)\Pi_{xy}^*\left[1-3e-\frac{1}{11\theta}(2-3e)\left(\Pi_{xx}^*+\Pi_{yy}^*\right)\right]$\\
	$\displaystyle \tilde{\Lambda}_{xy}^{(3)*}
	=\frac{1}{35}(1+e)^2$\\
	$\displaystyle \tilde{\Lambda}_{xy}^{(4)*}
	=\frac{1}{12012\sqrt{\pi}\theta}(1+e)\Pi_{xy}^*\left[19+15e-\frac{9}{2\theta}(1+e)\left(\Pi_{xx}^*+\Pi_{yy}^*\right)\right]$\\
	$\displaystyle \tilde{\Lambda}_{xy}^{(5)*}=0$\\
	$\displaystyle \tilde{\Lambda}_{xy}^{(6)*}
	 =-\frac{1}{583440\sqrt{\pi}\theta}(1+e)\Pi_{xy}^*\left[19+7e-\frac{35}{19\theta}(5+2e)\left(\Pi_{xx}^*+\Pi_{yy}^*\right)\right]$\\
	\hline
	$\displaystyle \delta\tilde{\Lambda}_{xx}^{(0)*}
	=\frac{24}{5\sqrt{\pi}}(1+e)(3-e)\left[\frac{1}{\theta}\Pi_{xx}^*+\frac{1}{42\theta^2}\left(\Pi_{xy}^{*2}+\Pi_{xx}^{*2}-2\Pi_{xx}^*\Pi_{yy}^*-2\Pi_{yy}^{*2}\right)\right]$\\
	$\displaystyle \delta\tilde{\Lambda}_{xx}^{(1)*}
	=-\frac{8}{105\theta}(1+e)(4-3e)\Pi_{xy}^*$\\
	$\begin{array}{ll}
 \delta\tilde{\Lambda}_{xx}^{(2)*}
	 =&\displaystyle{-\frac{4}{105\sqrt{\pi}}(1+e)\left\{1+3e-\frac{3}{2\theta}(3-e)\Pi_{xx}^*+\frac{2}{\theta}\Pi_{yy}^*+\frac{1}{22\theta^2}(2-3e)\Pi_{xy}^{*2}\right.}\\
	&\displaystyle{\left.+\frac{1}{132\theta^2}\left[(43-15e)\Pi_{xx}^{*2}+2(2-3e)\Pi_{xx}^*\Pi_{yy}^* - (35-3e)\Pi_{yy}^{*2}\right]\right\}}
\end{array}$\\
	$\displaystyle \delta\tilde{\Lambda}_{xx}^{(3)*}=0$\\
	$
\begin{array}{ll}
\delta\tilde{\Lambda}_{xx}^{(4)*}
	 =&\displaystyle{-\frac{1}{3465\sqrt{\pi}}(1+e)\left\{5+3e+\frac{1}{52\theta}(115-57e)\Pi_{xx}^*-\frac{1}{13\theta}(35+3e)\Pi_{yy}^*+\frac{45}{104\theta^2}(1+e)\Pi_{xy}^{*2}\right.}\nonumber\\
	 &\displaystyle{\left.-\frac{3}{104\theta^2}\left[(23-9e)\Pi_{xx}^{*2}-2(1+3e)\Pi_{xx}^*\Pi_{yy}^*-28\Pi_{yy}^{*2}\right]\right\}}
\end{array}
$\\
	$\displaystyle \delta\tilde{\Lambda}_{xx}^{(5)*}=0$\\
	$
\begin{array}{ll}
 \delta\tilde{\Lambda}_{xx}^{(6)*}
	=&\displaystyle{\frac{1}{360360\sqrt{\pi}}(1+e)\left\{3+e+\frac{3}{34\theta}(9-11e)\Pi_{xx}^* - \frac{6}{17\theta}(10+e)\Pi_{yy}^*+\frac{245}{646\theta^2}(5+2e)\Pi_{xy}^{*2}\right.}\\
	 &\displaystyle{\left.-\frac{5}{1292\theta^2}\left[(185-97e)\Pi_{xx}^{*2}-2(74+41e)\Pi_{xx}^*\Pi_{yy}^*-13(31+e)\Pi_{yy}^{*2}\right]\right\}}
\end{array}$\\
	\hline
	$\displaystyle \delta\tilde{\Lambda}_{yy}^{(0)*}
	=\frac{24}{5\sqrt{\pi}}(1+e)(3-e)\left[\frac{1}{\theta}\Pi_{yy}^*+\frac{1}{42\theta^2}\left(\Pi_{xy}^{*2}-2\Pi_{xx}^{*2}-2\Pi_{xx}^*\Pi_{yy}^*+\Pi_{yy}^{*2}\right)\right]$\\
	$\displaystyle \delta\tilde{\Lambda}_{yy}^{(1)*}
	=-\frac{8}{105\theta}(1+e)(4-3e)\Pi_{xy}^*$\\
	$
\begin{array}{ll}
 \delta\tilde{\Lambda}_{yy}^{(2)*}
	 =&\displaystyle{-\frac{4}{105\sqrt{\pi}}(1+e)\left\{1+3e+\frac{2}{\theta}\Pi_{xx}^*-\frac{3}{2\theta}(3-e)\Pi_{yy}^*+\frac{1}{22\theta^2}(2-3e)\Pi_{xy}^{*2}\right.}\\
	&\displaystyle{\left.-\frac{1}{132\theta^2}\left[(35-3e)\Pi_{xx}^{*2}-2(2-3e)\Pi_{xx}^*\Pi_{yy}^* - (43-15e)\Pi_{yy}^{*2}\right]\right\}}
\end{array}$\\
	$\displaystyle \delta\tilde{\Lambda}_{yy}^{(3)*}=0$\\
	$
\begin{array}{ll}
 \delta\tilde{\Lambda}_{yy}^{(4)*}
	 =&\displaystyle{-\frac{1}{3465\sqrt{\pi}}(1+e)\left\{5+3e-\frac{1}{13\theta}(35+3e)\Pi_{xx}^*+\frac{1}{52\theta}(115-57e)\Pi_{yy}^*+\frac{45}{104\theta^2}(1+e)\Pi_{xy}^{*2}\right.}\\
	 &\displaystyle{\left.+\frac{3}{104\theta^2}\left[28\Pi_{xx}^{*2}+2(1+3e)\Pi_{xx}^*\Pi_{yy}^*-(23-9e)\Pi_{yy}^{*2}\right]\right\}}
\end{array}$\\
	$\displaystyle \delta\tilde{\Lambda}_{yy}^{(5)*}=0$\\
	$
\begin{array}{ll}
 \delta\tilde{\Lambda}_{yy}^{(6)*}
	=&\displaystyle{\frac{1}{360360\sqrt{\pi}}(1+e)\left\{3+e - \frac{6}{17\theta}(10+e)\Pi_{xx}^*+\frac{3}{34\theta}(9-11e)\Pi_{yy}^*+\frac{245}{646\theta^2}(5+2e)\Pi_{xy}^{*2}\right.}\\
	 &\displaystyle{\left.+\frac{5}{1292\theta^2}\left[13(31+e)\Pi_{xx}^{*2}+2(74+41e)\Pi_{xx}^*\Pi_{yy}^*-(185-97e)\Pi_{yy}^{*2}\right]\right\}}
\end{array}$\\
	\end{tabular}
\end{ruledtabular}
\end{table}

\begin{table}
\begin{ruledtabular}
	\caption{Coefficients appearing in Eq.\ \eqref{eq:Pc_exp} up to order $N_{\text{c}}=6$. \label{fig:coeff2}}
 	\begin{tabular}{l}
 	 	$\displaystyle \tilde{\Pi}_{xy}^{c(0)*}
	=\frac{4}{5\theta}(1+e)\Pi_{xy}$\\
	$\displaystyle \tilde{\Pi}_{xy}^{c(1)*}
	=-\frac{4}{5\sqrt{\pi}}(1+e)\left[1+\frac{2}{7\theta}(\Pi_{xx}^*+\Pi_{yy})-\frac{1}{28\theta^2}\Pi_{xy}^{*2}
		-\frac{1}{84\theta^2}\left(7\Pi_{xx}^{*2}+9\Pi_{xx}^*\Pi_{yy}^*+7\Pi_{yy}^{*2}\right)\right]$\\
	$\displaystyle \tilde{\Pi}_{xy}^{c(2)*}=0$\\
	$\displaystyle \tilde{\Pi}_{xy}^{c(3)*}
	 =-\frac{1}{105\sqrt{\pi}}(1+e)\left[1-\frac{2}{11\theta}(\Pi_{xx}^*+\Pi_{yy}^*)+\frac{75}{572\theta^2}\Pi_{xy}^{*2}
		+\frac{3}{572\theta^2}\left(7\Pi_{xx}^{*2}+9\Pi_{xx}^*\Pi_{yy}^*+7\Pi_{yy}^{*2}\right)\right]$\\
	$\displaystyle \tilde{\Pi}_{xy}^{c(4)*}=0$\\
	$\displaystyle \tilde{\Pi}_{xy}^{c(5)*}
	 =\frac{1}{24024\sqrt{\pi}}(1+e)\left[1-\frac{3}{5\theta}(\Pi_{xx}^*+\Pi_{yy}^*)+\frac{49}{68\theta^2}\Pi_{xy}^{*2}
		+\frac{1}{68\theta^2}\left(13\Pi_{xx}^{*2}+19\Pi_{xx}^*\Pi_{yy}^*+13\Pi_{yy}^{*2}\right)\right]$\\
	$\displaystyle \tilde{\Pi}_{xy}^{c(6)*} = 0$\\
	\end{tabular}
\end{ruledtabular}
	\end{table}

By expanding the right-hand sides of Eqs.\ \eqref{I2-I3-Jx-Jy} in powers of $b_T$, 
the angle integrals in Eqs.\ \eqref{eq:Pc*-Lambda*} and \eqref{Lab*-Mab*} can be analytically performed term by term. 
This gives rise to an expansion of the quantities $\Lambda_{\alpha\alpha}^*$, 
$\Lambda_{xy}^*$, $\delta\Lambda_{xx}^*$,  $\delta\Lambda_{yy}^*$, and $\Pi_{xy}^{c*}$ in powers of the dimensionless parameter
\begin{equation}
  \tilde{\dot{\gamma}}\equiv\frac{\dot\gamma^*}{\xi_{\rm ex}\sqrt{\theta}}
\end{equation}
as
\begin{subequations}
\label{eq:Lii_exp-eq:Lyy_exp}
\begin{align}
	\Lambda_{\alpha\alpha}^*
	&= \varphi g_0 \xi_{\rm ex}\theta^{3/2}\sum_{i=0}^{N_{\text{c}}} \tilde{\Lambda}_{\alpha\alpha}^{(i)*}\tilde{\dot{\gamma}}^i,\label{eq:Lii_exp}\\
	\Lambda_{xy}^*&= \varphi g_0 \xi_{\rm ex}\theta^{3/2}\sum_{i=0}^{N_{\text{c}}} \tilde{\Lambda}_{xy}^{(i)*}\tilde{\dot{\gamma}}^i,\\
	\delta\Lambda_{xx}^*&= \varphi g_0 \xi_{\rm ex}\theta^{3/2}\sum_{i=0}^{N_{\text{c}}} \delta\tilde{\Lambda}_{xx}^{(i)*}\tilde{\dot{\gamma}}^i,\\
	\delta\Lambda_{yy}^*&= \varphi g_0 \xi_{\rm ex}\theta^{3/2}\sum_{i=0}^{N_{\text{c}}} \delta\tilde{\Lambda}_{yy}^{(i)*}\tilde{\dot{\gamma}}^i,\label{eq:Lyy_exp}\\
	\Pi_{xy}^{c*}&=\varphi g_0 \theta\sum_{i=0}^{N_{\text{c}}} \tilde\Pi_{xy}^{c(i)*}\tilde{\dot{\gamma}}^i. \label{eq:Pc_exp}
\end{align}
\end{subequations}
Here, we have introduced an upper cutoff $N_{\text{c}}$ in the series for practical  reasons.
Equations \eqref{eq:Lii_exp-eq:Lyy_exp} are exact within the framework of the Enskog approximation and Grad's expansion if we take the limit $N_{\text{c}}\to\infty$. 
This is equivalent to the closed forms obtained from Eqs.\ \eqref{eq:Pc*-Lambda*}, \eqref{Lab*-Mab*}, and \eqref{I2-I3-Jx-Jy}.
From a  practical point of view, however, it is computationally much more convenient  to introduce a truncation up to a finite number of terms (finite $N_{\text{c}}$).
The coefficients in Eqs.~\eqref{eq:Lii_exp}--\eqref{eq:Lyy_exp} and in Eq.~\eqref{eq:Pc_exp} up to sixth order in the shear rate are listed in Tables \ref{fig:coeff} and \ref{fig:coeff2}, respectively. 
Truncating up to $N_{\text{c}}=1$, as well as  neglecting quadratic terms in $\Pi_{\alpha\beta}$, yields results consistent with those derived in Ref.~\cite{Hayakawa17} within the linear shear rate approximation.
It is also worthwhile noting that $\tilde{\dot{\gamma}}\propto \varphi\dot{\gamma}/\nu$, where the collision frequency $\nu$ is defined by 
\begin{equation}
	\nu=\frac{2\sqrt{2\pi}}{5} (1+e)(3-e)n \sigma^{2}v_T.
\end{equation}
Therefore, in the low-density regime $\varphi\to 0$ (Boltzmann limit), only the terms with $i=0$ in Eqs.\ \eqref{eq:Lii_exp-eq:Lyy_exp} survive.

\begin{figure}[htbp]
\includegraphics[width=\linewidth]{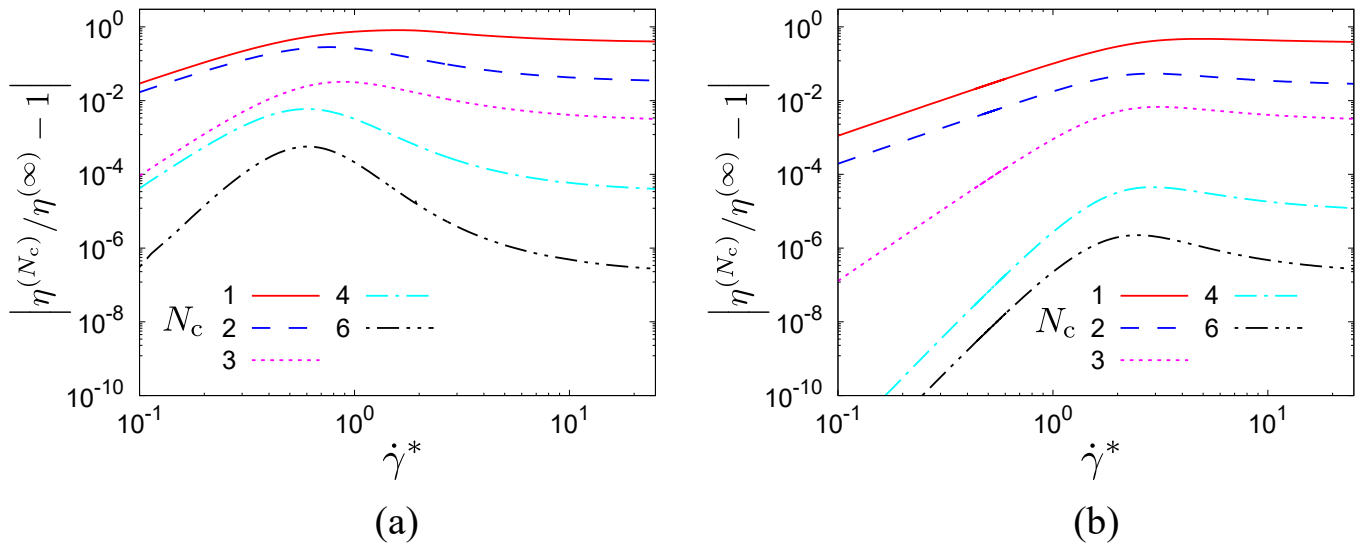}
\caption{
Convergence of the shear viscosity against the truncation order in Eqs.\ \eqref{eq:Lii_exp-eq:Lyy_exp} for  $\varphi=0.30$ and  $e=0.9$. Two values of $\xi_{\text{ex}}$ are considered: (a) $\xi_{\text{ex}}=0.1$ and (b) $\xi_{\text{ex}}=1.0$.
Here, $\eta^{(N_{\text{c}})}$ represents the shear viscosity with the truncation up to  order $N_{\text{c}}$, while $\eta^{(\infty)}$ is the non-truncated viscosity.
}
\label{fig:vis}
\end{figure}

Figure \ref{fig:vis} shows the convergence of the stationary viscosity depending on the truncation order for $\varphi=0.30$, $e=0.9$, and two values of the parameter $\xi_{\text{ex}}$, namely  $\xi_{\text{ex}}=0.1$ and $\xi_{\text{ex}}=1.0$.
We can observe that the expansions  in Eqs.~\eqref{eq:Lii_exp-eq:Lyy_exp} have a rather fast convergence.
The maximum relative deviations from the non-truncated values in the cases with $\xi_{\text{ex}}=0.1$ ($\xi_{\text{ex}}=1$) are observed to be $0.81$ ($0.47$), $0.28$ ($0.054$), $0.032$ ($6.7\times 10^{-3}$), $5.9\times 10^{-3}$ ($4.5\times 10^{-5}$), and $5.7\times 10^{-4}$ ($2.2\times 10^{-6}$) for $N_{\text{c}}=1$, $2$, $3$, $4$, and $6$, respectively.
Since the perturbation parameter $\tilde{\dot{\gamma}}$ used in Eqs.\ \eqref{eq:Lii_exp-eq:Lyy_exp} is inversely proportional to $\xi_{\text{ex}}$, 
it is not surprising that the convergence is much better in the case $\xi_{\text{ex}}=1.0$ than in the case $\xi_{\text{ex}}=0.1$. Additionally, as will be shown in Sec.\  \ref{sec:simulation}, 
the steady-state  dimensionless temperature $\theta$ increases monotonically with $\dot{\gamma}^*$ and, as a consequence, $\tilde{\dot{\gamma}}$ exhibits a nonmonotonic dependence on $\dot{\gamma}^*$. 
More specifically, we have observed (not shown) that $\tilde{\dot{\gamma}}$ has a maximum value $\tilde{\dot{\gamma}}=3.57$ ($\tilde{\dot{\gamma}}=1.42$) at $\dot{\gamma}^*\simeq 0.58$ ($\dot{\gamma}^*\simeq 2.4$) if $\xi_{\text{ex}}=0.1$ ($\xi_{\text{ex}}=1.0$). 
This nonmonotonic dependence of $\tilde{\dot{\gamma}}$  on $\dot{\gamma}^*$ explains the nonmonotonic behavior of the relative errors of the truncated approximations observed in Fig.\ \ref{fig:vis}. 
Since the error of the sixth-order approximation is less than $0.06\%$ and $0.0003\%$ for $\xi_{\text{ex}}=0.1$ and $\xi_{\text{ex}}=1.0$, respectively, all the theoretical results presented in Sec.\ \ref{sec:simulation} have been obtained with the choice $N_{\text{c}}=6$.

\section{Comparison between theory and simulation}\label{sec:simulation}

\begin{figure}[htbp]
\includegraphics[width=\linewidth]{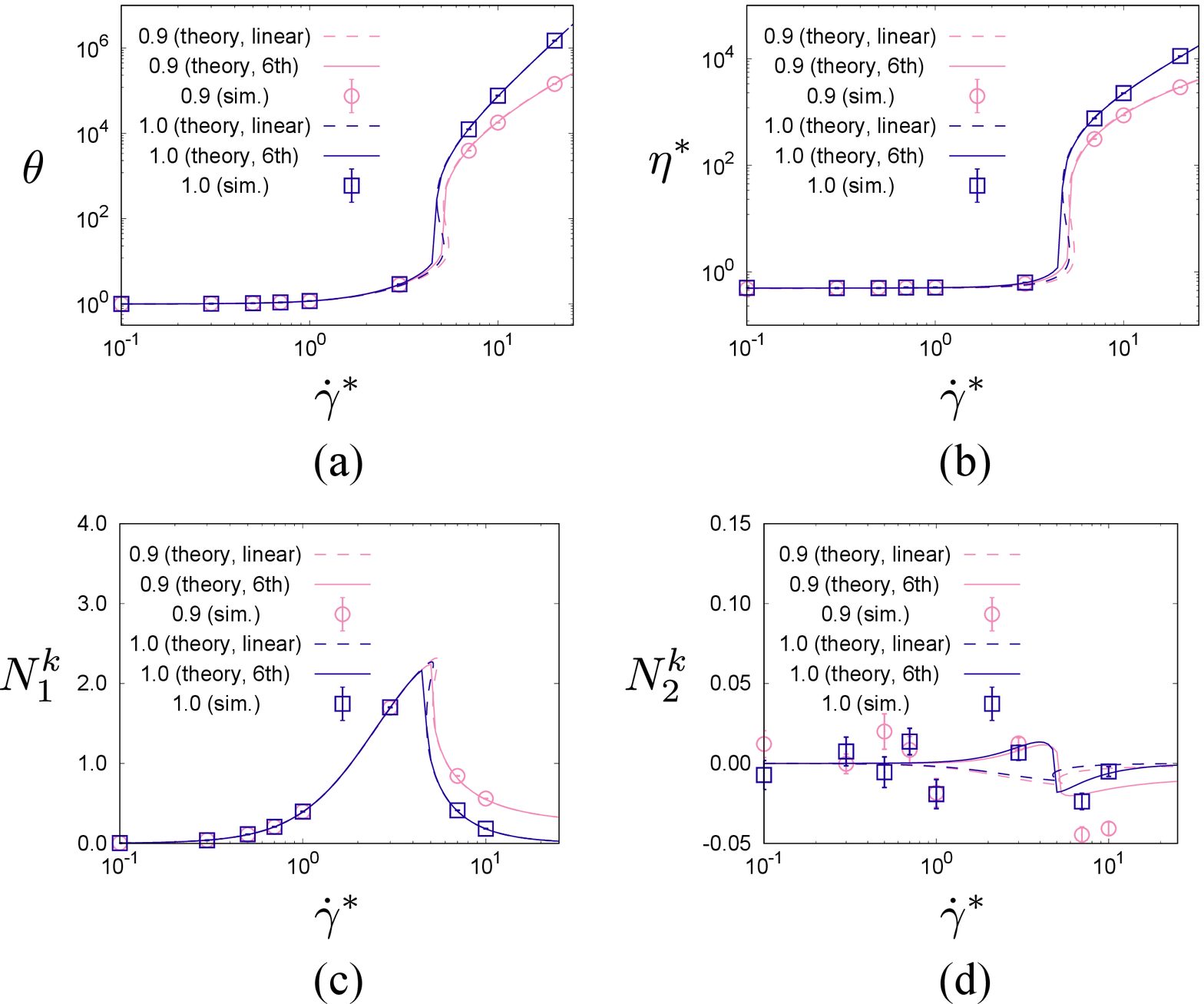}
\caption{
Plots of (a) $\theta$, (b) $\eta^*$, (c) $N_1^k$, and (d) $N_2^k$ versus the dimensionless shear rate $\dot\gamma^{*}$ for $\varphi=0.01$ and two different values of the restitution coefficient: $e=1$ and $e=0.9$. The solid and dashed lines correspond to the (perturbative) theoretical results obtained in the sixth-order (denoted by ``6th'' in the legend) and first-order (denoted by ``linear'' in the legend), respectively. Symbols refer to computer simulation results.}
\label{fig1}
\end{figure}

\begin{figure}[htbp]
\includegraphics[width=\linewidth]{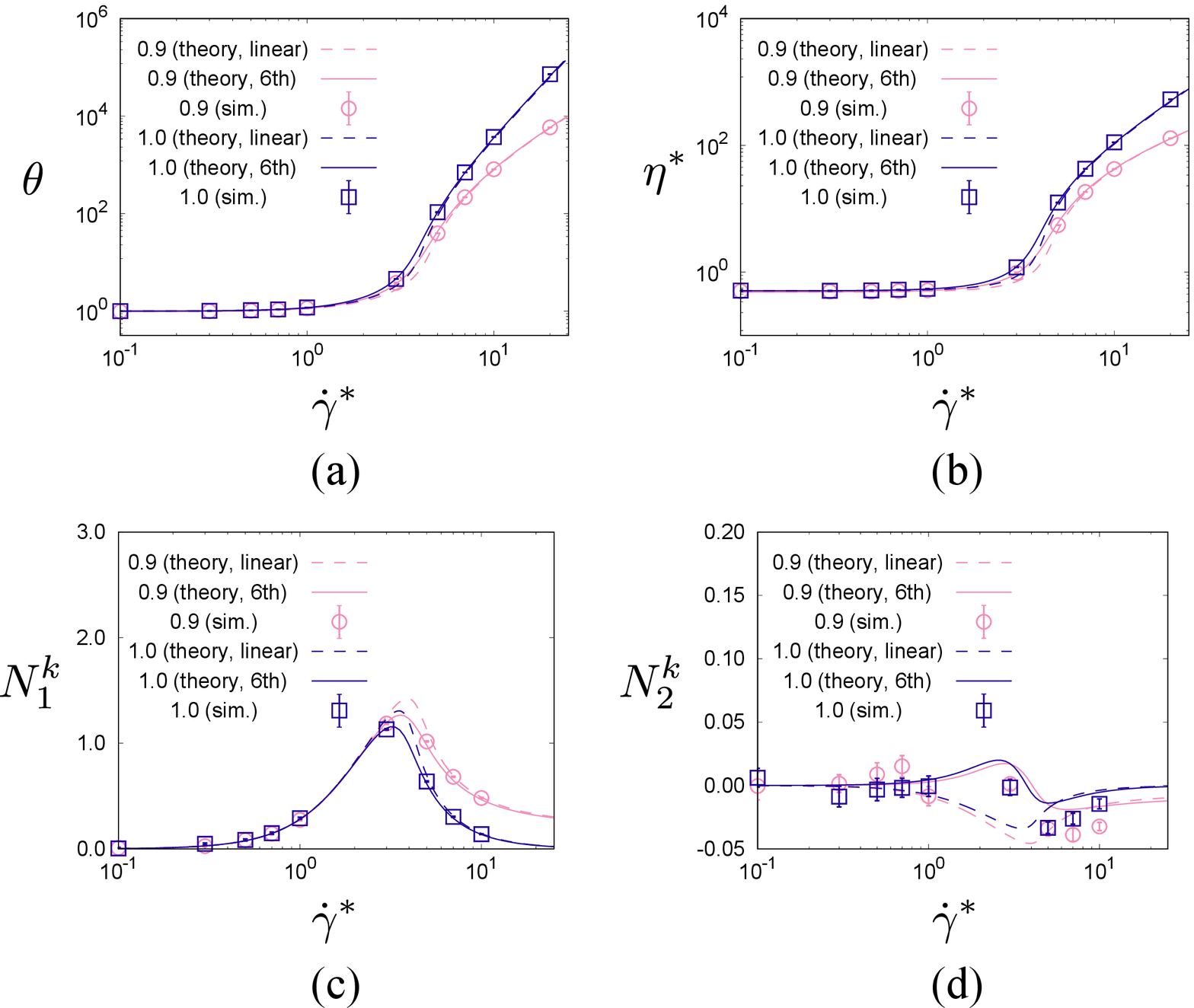}
\caption{
Plots of (a) $\theta$, (b) $\eta^*$, (c) $N_1^k$, and (d) $N_2^k$ versus the dimensionless shear rate $\dot\gamma^{*}$ for $\varphi=0.05$ and two different values of the restitution coefficient: $e=1$ and $e=0.9$. The solid and dashed lines correspond to the (perturbative) theoretical results obtained in the sixth-order (denoted by ``6th'' in the legend) and first-order (denoted by ``linear'' in the legend), respectively. Symbols refer to computer simulation results.}
\label{fig2}
\end{figure}

\begin{figure}[htbp]
\includegraphics[width=\linewidth]{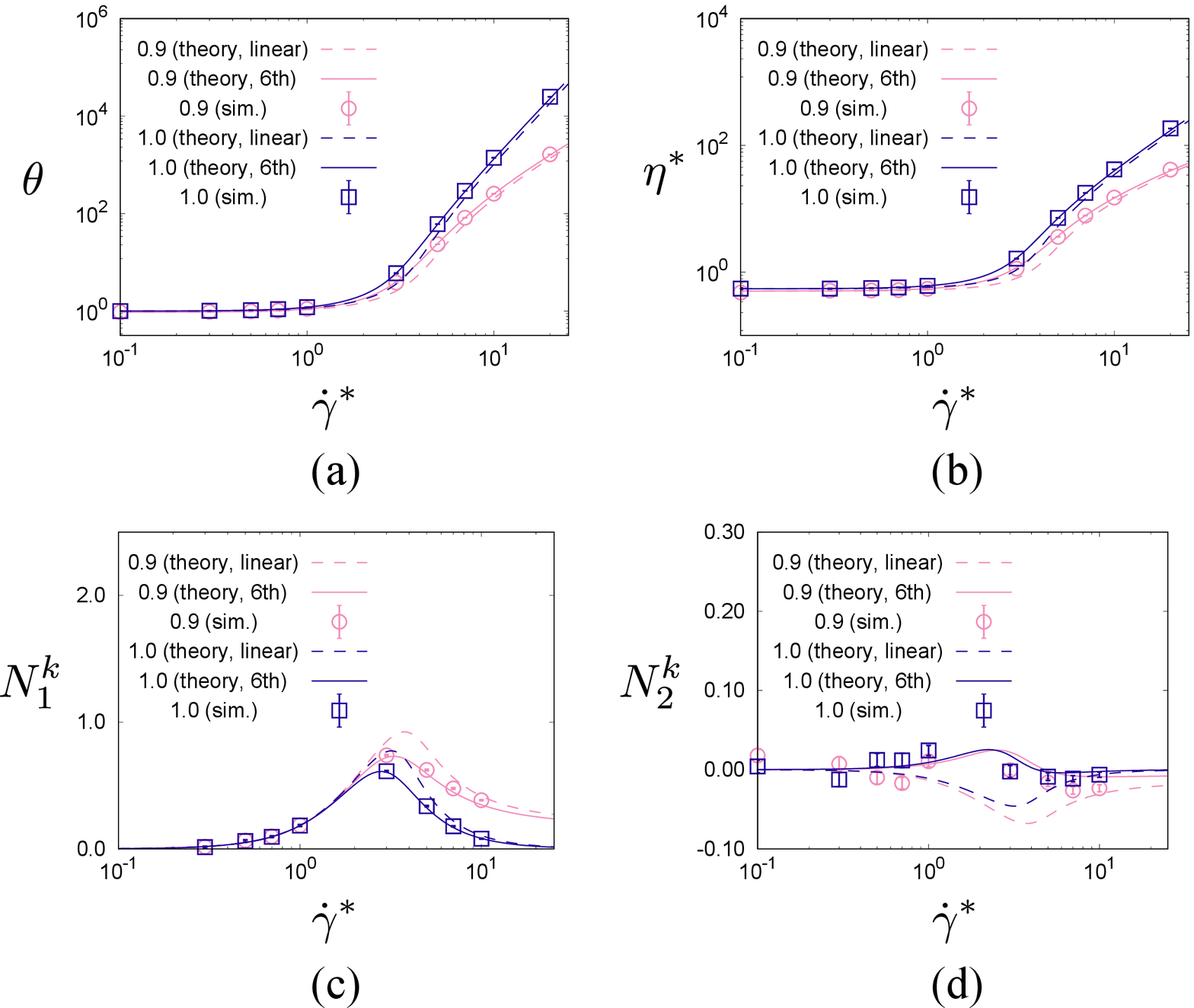}
\caption{
Plots of (a) $\theta$, (b) $\eta^*$, (c) $N_1^k$, and (d) $N_2^k$ versus the dimensionless shear rate $\dot\gamma^{*}$ for $\varphi=0.10$ and two different values of the restitution coefficient: $e=1$ and $e=0.9$. The solid and dashed lines correspond to the (perturbative) theoretical results obtained in the sixth-order (denoted by ``6th'' in the legend) and first-order (denoted by ``linear'' in the legend), respectively. Symbols refer to computer simulation results.}
\label{fig3}
\end{figure}

\begin{figure}[htbp]
\includegraphics[width=\linewidth]{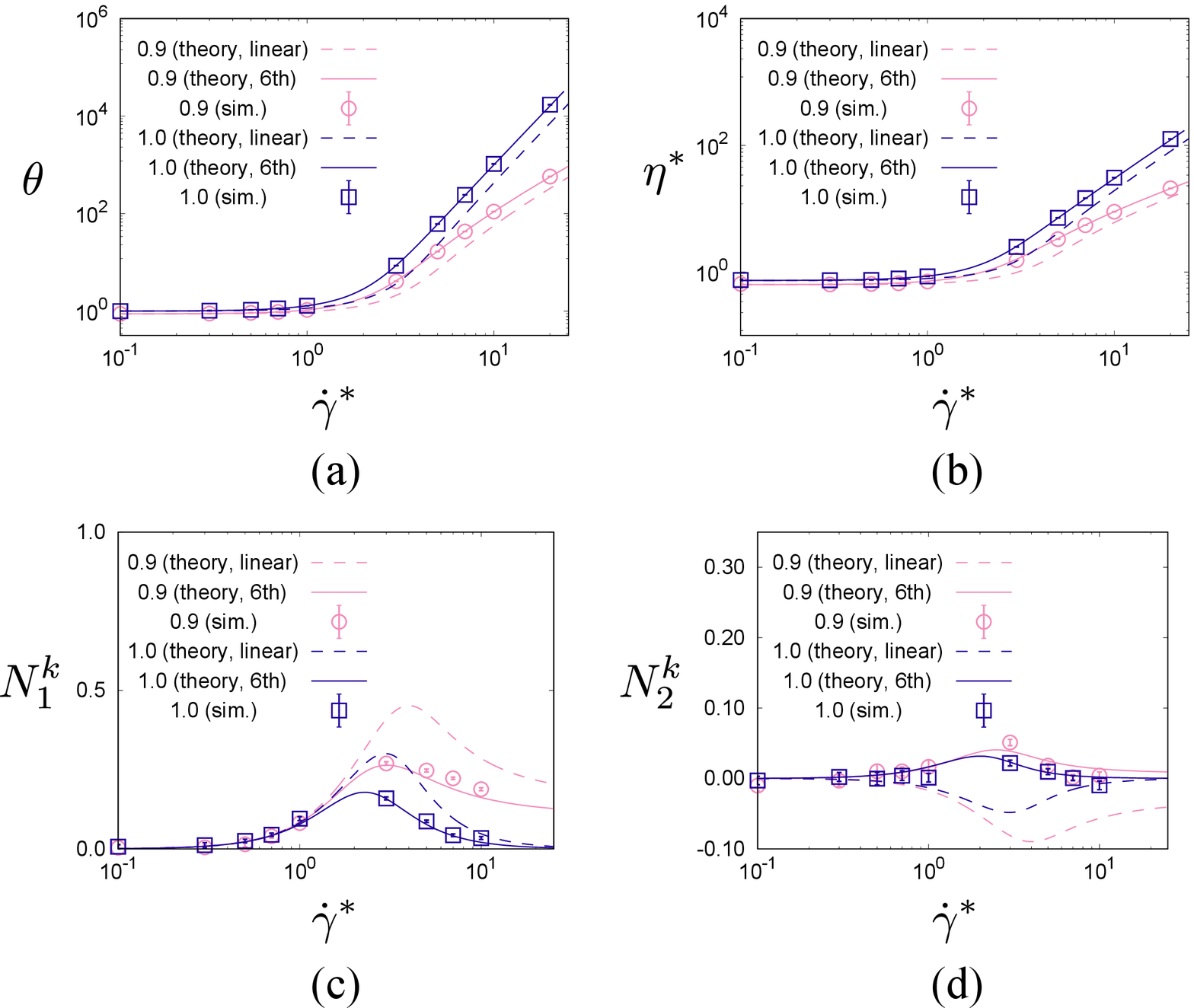}
\caption{
Plots of (a) $\theta$, (b) $\eta^*$, (c) $N_1^k$, and (d) $N_2^k$ versus the dimensionless shear rate $\dot\gamma^{*}$ for $\varphi=0.20$ and two different values of the restitution coefficient: $e=1$ and $e=0.9$. The solid and dashed lines correspond to the (perturbative) theoretical results obtained in the sixth-order (denoted by ``6th'' in the legend) and first-order (denoted by ``linear'' in the legend), respectively. Symbols refer to computer simulation results.}
\label{fig4}
\end{figure}

\begin{figure}[htbp]
\includegraphics[width=\linewidth]{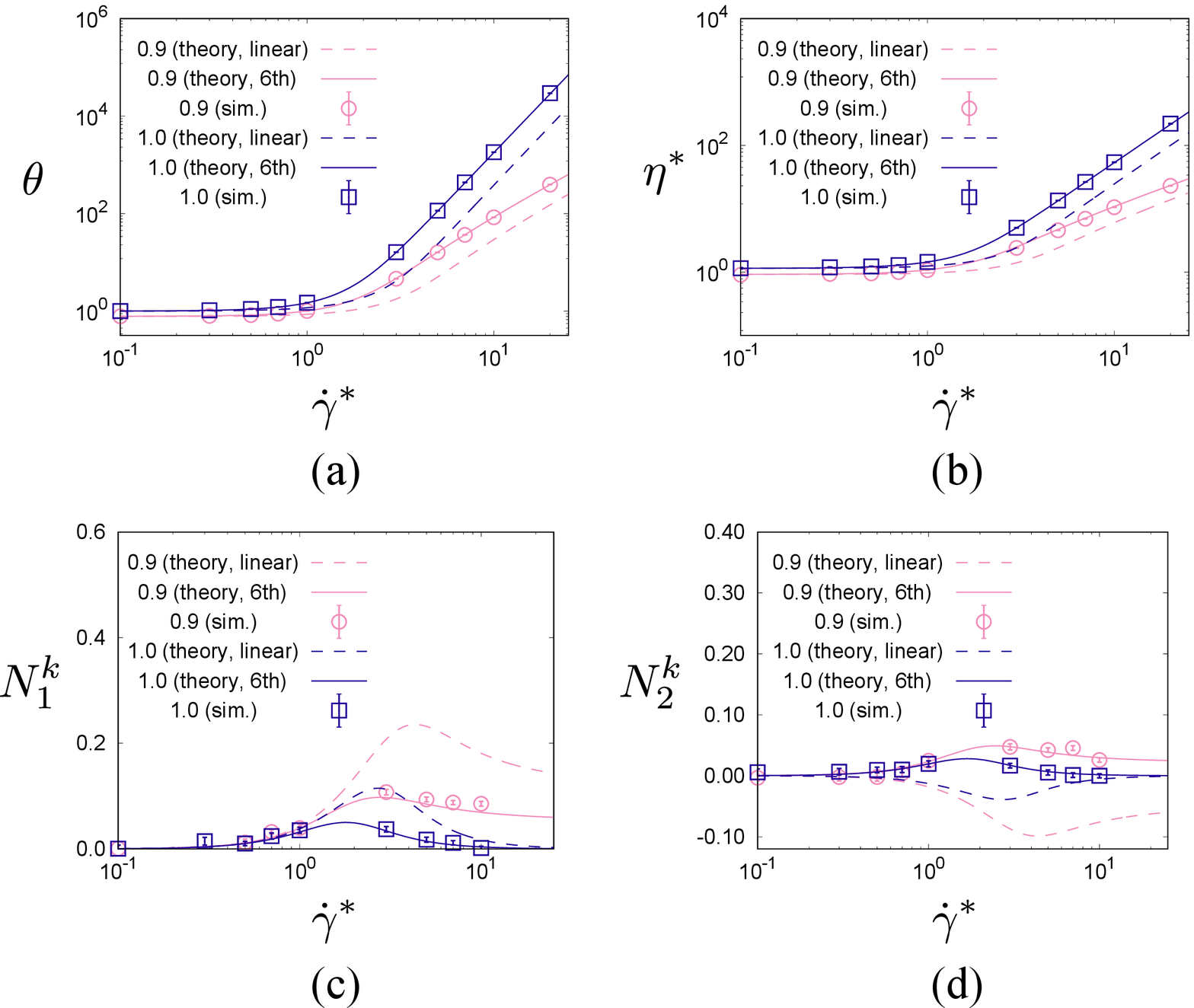}
\caption{
Plots of (a) $\theta$, (b) $\eta^*$, (c) $N_1^k$, and (d) $N_2^k$ versus the dimensionless shear rate $\dot\gamma^{*}$ for $\varphi=0.30$ and two different values of the restitution coefficient: $e=1$ and $e=0.9$. The solid and dashed lines correspond to the (perturbative) theoretical results obtained in the sixth-order (denoted by ``6th'' in the legend) and first-order (denoted by ``linear'' in the legend), respectively. Symbols refer to computer simulation results.}
\label{fig5}
\end{figure}

\begin{figure}[htbp]
\includegraphics[width=\linewidth]{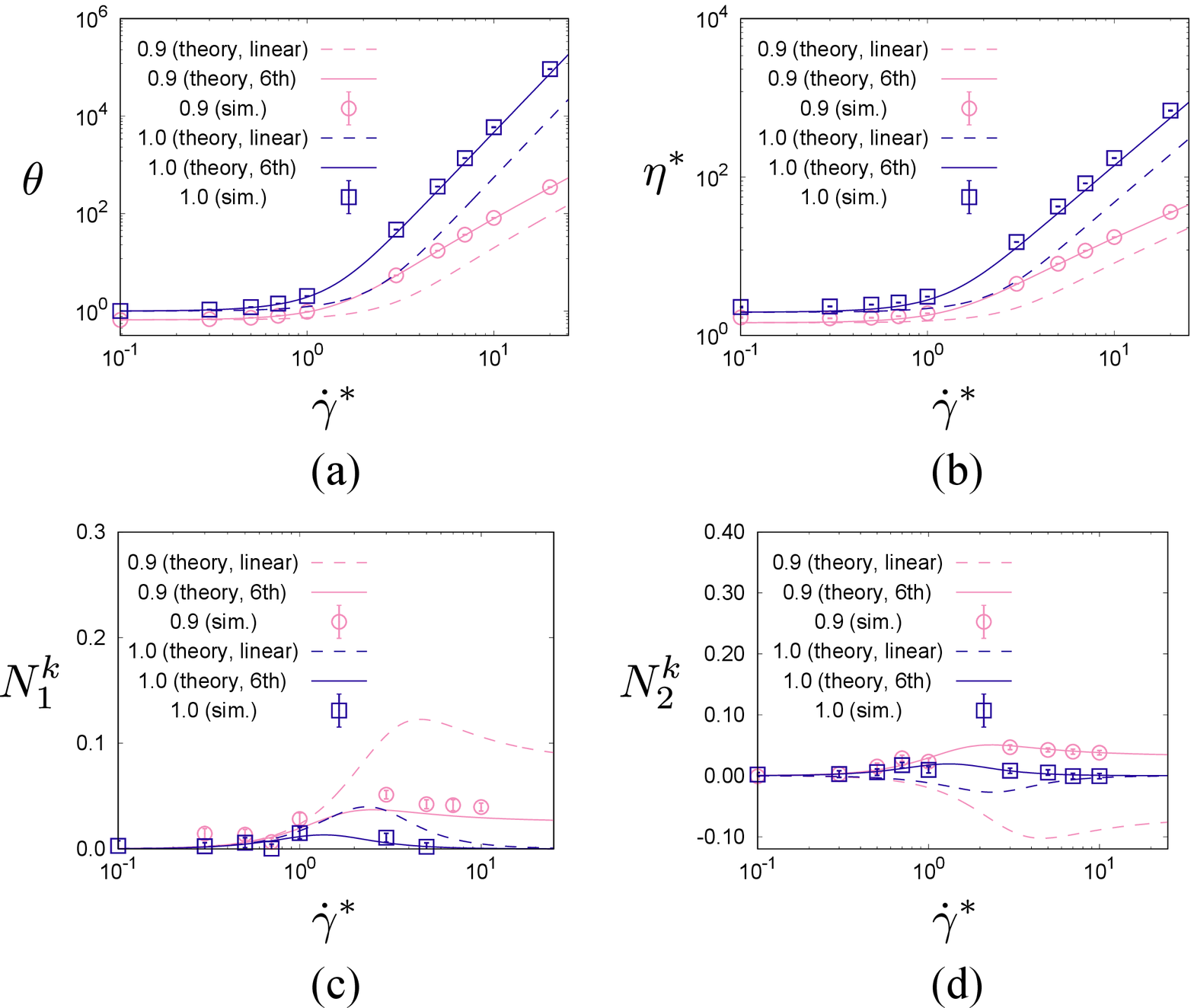}
\caption
{
Plots of (a) $\theta$, (b) $\eta^*$, (c) $N_1^k$, and (d) $N_2^k$ versus the dimensionless shear rate $\dot\gamma^{*}$ for $\varphi=0.40$ and two different values of the restitution coefficient: $e=1$ and $e=0.9$. 
The solid and dashed lines correspond to the (perturbative) theoretical results obtained in the sixth-order (denoted by ``6th'' in the legend) and first-order (denoted by ``linear'' in the legend), respectively. Symbols refer to computer simulation results.
}
\label{fig6}
\end{figure}

\begin{figure}[htbp]
\includegraphics[width=\linewidth]{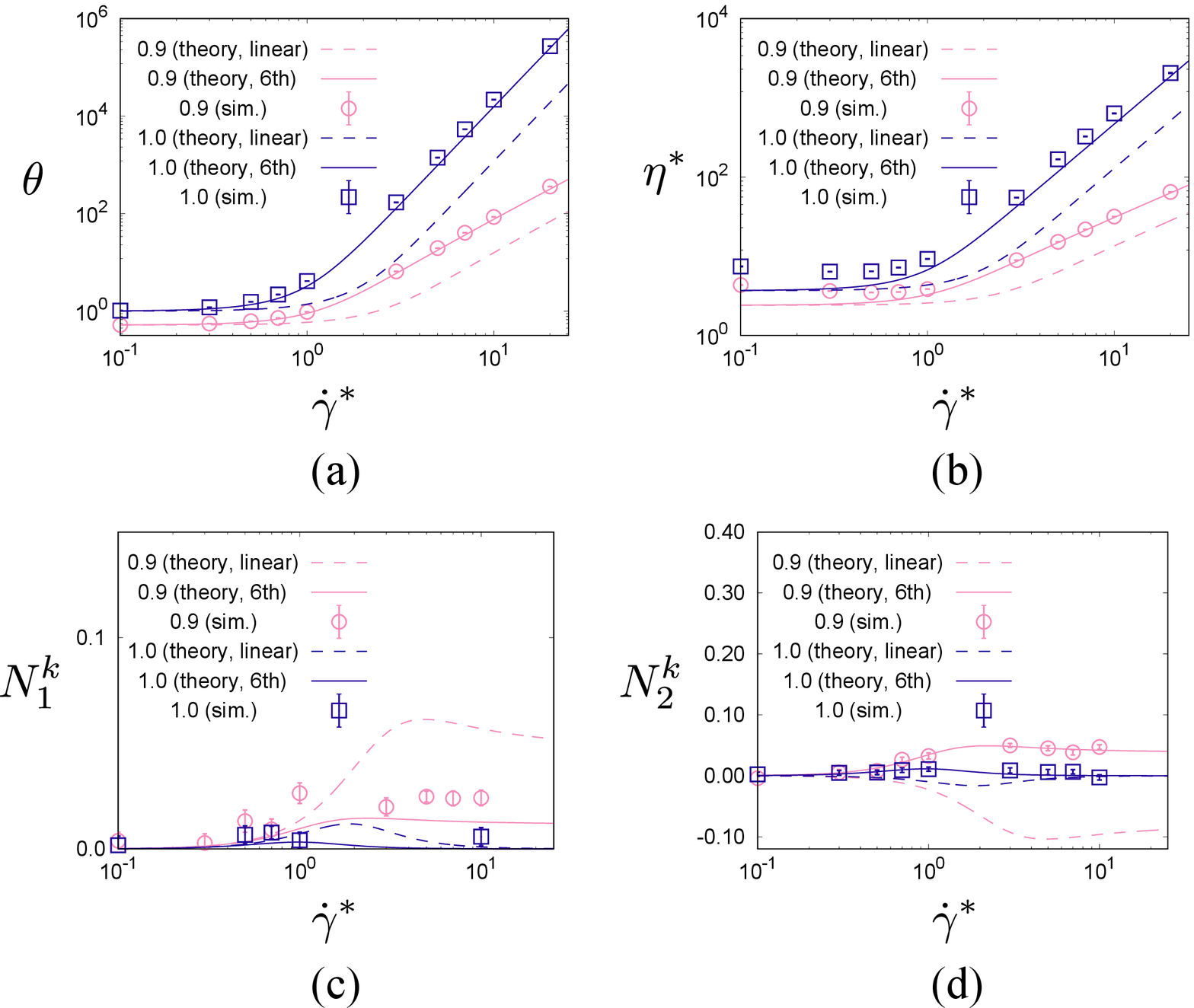}
\caption{
Plots of (a) $\theta$, (b) $\eta^*$, (c) $N_1^k$, and (d) $N_2^k$ versus the dimensionless shear rate $\dot\gamma^{*}$ for $\varphi=0.50$ and two different values of the restitution coefficient: $e=1$ and $e=0.9$. The solid and dashed lines correspond to the (perturbative) theoretical results obtained in the sixth-order (denoted by ``6th'' in the legend) and first-order (denoted by ``linear'' in the legend), respectively. Symbols refer to computer simulation results.}
\label{fig7}
\end{figure}

The goal of this section is to validate our theoretical results by using  the EDLSHS method.
We adopt Lees--Edwards boundary conditions in a three-dimensional periodic box~\cite{LE72,Scala12}.
Under these conditions, the Langevin equation \eqref{Langevin_eq} is equivalent to Eqs.~\eqref{Enskog} and \eqref{J(V|f)}, when molecular chaos ansatz and the Enskog approximation are assumed.

Notice that it is difficult to adopt either the conventional event-driven  or the soft-core simulation methods for our problem. 
The existence of both the inertia term $d\bm{p}/dt$ and the drag term proportional to $\zeta$ in Eq.~\eqref{Langevin_eq} makes it difficult the use of  conventional event-driven simulations. 
In addition, a sudden increment of the viscosity in the vicinity of a DST gives rise to numerical difficulties in soft-core simulations.
Thus, to avoid the above difficulties, we adopt EDLSHS \cite{Scala12}. This is in fact a powerful simulator for hard spheres under the influence of
the drag and the inertia terms with the aid of Trotter decomposition~\cite{Scala12,DST16}.

In our simulations, we fix the number of grains $N=1000$, as well as the background fluid temperature characterized by $\xi_{\rm ex}=1.0$. 
Several volume fractions  are considered: $\varphi=0.01$, $0.05$, $0.10$, $0.20$, $0.30$, $0.40$, and $0.50$. 
The first density corresponds to a dilute suspension, while the latter can be considered as a relatively dense suspension.
Notice that previous works \cite{LBD02,DHGD02,MGAL06,MDCPH11,MGH14} have shown that the results derived from the Enskog equation are quite accurate for moderately dense systems.
Two different values of the restitution coefficient $e$ are considered in this section: $e=1$ (elastic grains) and $e=0.9$ (granular grains with moderate inelasticity).
All the rheological variables presented in this paper are measured after the system reaches a steady state (for $t>5/\zeta$).
In addition, all the variables are averaged over $10$ ensemble averages, which have different initial conditions, and $10$ time averages during the time intervals $5/\zeta$ for each initial condition.
We have confirmed that the fluctuations of the observables are sufficiently small.

Figures \ref{fig1}--\ref{fig7} show the shear-rate dependence of the dimensionless kinetic temperature $\theta$, the apparent shear viscosity $\eta^*\equiv -(\Pi_{xy}^*+\Pi_{xy}^{c*})/\dot\gamma^*$, and the viscometric quantities
\begin{subequations}
\begin{align}\label{N_1^k}
N_1^k \equiv& \frac{P_{xx}^k-P_{yy}^k}{nT}=\frac{\Delta\theta}{\theta},\\
\label{N_2^k}
 N_2^k \equiv& \frac{P_{yy}^k-P_{zz}^k}{nT}
 =\frac{\delta\theta-\Delta\theta}{\theta},
 \end{align}
 \end{subequations}
for $\varphi=0.01$ (Fig.\ \ref{fig1}), $\varphi=0.05$ (Fig.\ \ref{fig2}), $\varphi=0.10$ (Fig.\ \ref{fig3}), $\varphi=0.20$ (Fig.\ \ref{fig4}), $\varphi=0.30$ (Fig.\ \ref{fig5}),  $\varphi=0.40$ (Fig.\ \ref{fig6}), and $\varphi=0.50$ (Fig.\ \ref{fig7}).
The dashed lines in those plots correspond to the theoretical results obtained by retaining the first-order shear rate as explained in Ref.~\cite{Hayakawa17} within the linear shear rate approximation.
These results will be referred here to as the first-order theory.
Analogously, the solid lines refer to the theoretical results by using the sixth-order expansion explained in Sec.\ \ref{sec:arbitrary_shear}.
The symbols in Figs.\ \ref{fig1}--\ref{fig7} correspond to the simulation results.

It is remarkable that, for $\varphi\le 0.4$, an excellent agreement is found  between  the results of our simulation for $\theta$, $\eta^*$, and $N_1^k$ and the theoretical results if we adopt the sixth-order expansions ($N_{\rm c}=6$) in Eqs.~\eqref{eq:Lii_exp-eq:Lyy_exp}, 
together  with the expressions in Tables \ref{fig:coeff} and \ref{fig:coeff2}.  This good agreement is shown more in detail in Table \ref{fig:comp_eta} in the case of the viscosity for $\varphi=0.10$ and $e=0.9$. 
Even at $\varphi=0.5$ (slightly above the Alder transition point $\varphi=0.49$), the sixth-order theory performs reasonably well (see Fig.\ \ref{fig7}), especially in the case of $\theta$.

The second viscometric function $N_2^k$ is rather small in the more dilute cases (see Figs.\ \ref{fig1} and \ref{fig2}). In fact, $N_2^k\to 0$ if $\varphi\to 0$ in Grad's approximation when the terms nonlinear in  $\Pi_{\alpha\beta}$ are neglected; the analysis of such nonlinear contributions for dilute cases can be found in Ref.~\cite{DST16}.
As a consequence, the agreement between simulation and theory in panels (d) of Figs.\ \ref{fig1}--\ref{fig3} is worse than in panels (d) of Figs.\ \ref{fig4}--\ref{fig7}.
In other words, our theory gives precise results of $N_2^k$ for $\varphi\ge 0.2$.
It is interesting to note that both $N_1^k$ and $N_2^k$ have peaks at around  the bending points of $\theta$ and $\eta^*$, their peak values being enhanced if the  collisions are inelastic.

The first-order theory also gives reasonable results for $\varphi\le 0.1$, in which case the high shear-rate contributions are dominated by those of the dilute theory. As density increases, however, the first-order theory becomes less reliable.

\begin{table}
	\caption{Comparison of the shear viscosity between the simulation and the sixth-order theory for $\varphi=0.10$ and $e=0.9$.}
 	\begin{ruledtabular}
 \begin{tabular}{cccc}
 	 	 & simulation & theory & relative deviation\\
 	$\dot\gamma$ & $\eta^*_{\rm sim}$ & $\eta^*_{\rm th}$ & $|\eta^*_{\rm sim}-\eta^*_{\rm th}|/\eta^*_{\rm sim}$ \\
 	\hline
 	$0.50$ & $0.512899$ & $0.513355$ & $8.88\times10^{-4}$ \\
 	$1.0$  & $0.544695$ & $0.547774$ & $5.65\times10^{-3}$ \\
 	$3.0$  & $1.13576$  & $1.18055$ & $3.94\times10^{-2}$ \\
 	$5.0$  & $3.60394$  & $3.73449$ & $3.62\times10^{-2}$ \\
 	$7.0$  & $7.79684$  & $7.80020$ & $4.31\times10^{-4}$ \\
 	$10.0$ & $15.0416$ & $14.7821$ & $1.73\times10^{-2}$ \\
		\end{tabular}
	\label{fig:comp_eta}
\end{ruledtabular}
\end{table}

\section{Discussion and conclusion}\label{sec:discussion}
The Enskog kinetic equation for inelastic hard spheres has been considered in this paper as the starting point to study the rheology of inertial suspensions under simple shear flow.
The effect of the interstitial fluid on the dynamics of solid particles has been modeled through a viscous drag force plus a stochastic Langevin-like term.
While the first term models the friction of grains on the continuous phase, the latter accounts for thermal fluctuations.
Two independent but complementary routes have been employed to determine the non-Newtonian transport properties of the suspended particles.
First, the Enskog equation has been approximately solved by means of Grad's moment method.
Then, the theoretical results for the kinetic temperature, the viscosity, and the first and second normal stress differences have been compared against computer simulations based on the event-driven Langevin simulation for hard spheres (EDLSHS) \cite{Scala12}.
The main goal of the paper has been to study the influence of  both inelasticity and density (or volume fraction) on the flow curve (stress-strain rate relation).

The analysis in this paper includes nonlinear effects in the shear rate, thus overcoming the limitations of the linear theory presented in Ref.~\cite{Hayakawa17}.
As a result, the theoretical results derived in this paper from Grad's method indicate that the Enskog theory describes well the rheology of sheared suspensions.
In particular, the agreement found between theory and simulations for the shear viscosity clearly shows that the shear thickening effect is well captured by the Enskog kinetic equation in combination with Grad's method.
Our analysis can be regarded as the complete version of previous works~\cite{DST16,Tsao95,BGK2016,Sangani96,Chamorro15,Saha17,Hayakawa17}, some of which only discuss the transition between the quenched state and the ignited state for the kinetic temperature \cite{Tsao95,Sangani96,Saha17}.
We have to stress that the theory predicts precise results without any fitting parameters and  has a wide applicability  for $\varphi\le 0.5$, at least for $\theta$, and for $\varphi\le 0.4$ for $\eta$.
This confirms the reliability of the Enskog equation in this range of densities reported in previous works \cite{LBD02,DHGD02,MGAL06,MDCPH11,MGH14}.

Typical DSTs observed in experiments and simulations for dense suspensions ($\varphi>0.5$) are essentially the result of mutual friction between grains.
Although the Enskog kinetic equation for hard spheres is not applicable to such dense suspensions, an extension of Grad's moment method to dense soft systems for frictionless grains~\cite{Suzuki17} might be applicable for the explanation of the DST in frictional grains, thus improving over a previous theory of dense granular liquids~\cite{Suzuki15}.
This study will be reported elsewhere~\cite{Saitoh17} (see also Ref.~\cite{Saitoh16}).
We should note that contact states between grains are important to describe typical DSTs in dense suspensions, which cannot be included in the model of hard spheres because it is described by instantaneous collisions.
Therefore, we must model the process in terms of a soft-core model having finite duration time.

The Langevin equation \eqref{Langevin_eq} employed in our study assumes that the gravity force is perfectly balanced with the drag force associated with the fluid flow.
 This assumption is only true if the homogeneous state is stable.
On the other hand, the simple shear flow state becomes unstable above a critical shear rate.
If the homogeneous state is unstable, one would need to consider the time evolution of local structures as well as the consideration of an inhomogeneous drag.

The assumption $e={\rm const}$ has allowed us to achieve explicit results. 
However, experimental observations \cite{BHL84} as well as the mechanics of particle collisions \cite{RPBS99}  showed that the restitution coefficient $e$ must be a function of the impact velocity. 
One of the simplest models accounting for the velocity dependence of $e$ is that of viscoelastic particles \cite{BP00,BP03,DBPB13}, 
for which some progresses have been made in the case of quasielastic particles. 
However, the extension of the present results to a model with a velocity-dependent restitution coefficient is beyond the scope of this paper. 
In any case, as already pointed out in Ref.\ \cite{Hayakawa17}, given that the transition between DST and CST for elastic suspensions is qualitatively similar to that for inelastic suspensions (except in the high-shear asymptotic region), 
we believe that the effect of the velocity dependence of the restitution coefficient is not especially important in the shear thickening problem.

Our Enskog kinetic theory assumes that there is no long-time correlation in each collision. 
However, it is known that the noise term generates long-time tails in the treatment of fluctuating hydrodynamics.
Although we do not have any definite answer to the role of such long tails,
we have some comments on the tails.
(i) Basically, the effect of long-time tail is harmless in 3D cases.
(ii)	It is known that the correction from long-time tail is almost invisible for the 3D case because its prefactor is too small, as indicated by the classical work~\cite{Yamada75}.
(iii)	The long tails are suppressed in sheared fluids because the slope of the tail becomes steep for $t>\dot\gamma^{-1}$.
(iv)	A simple Green--Kubo formula in which the time integral of the stress-stress time correlation gives the viscosity may not be used in highly sheared granular fluids. 
(v)	Such effects might be absorbed by Grad's approximation, because the application of the Green--Kubo formula including the effect of long tails predicts small CST.
It would be interesting to clarify the reason why such long correlation effects (both time and spatial)~\cite{Otsuki09a,Otsuki09b} are irrelavant in inertial suspensions.

Finally, it is worthwhile noting that the monodisperse system studied in this paper crystallizes in the case of volume fractions larger than $0.49$ for low shear rates. This crystallization could be prevented by considering sheared polydisperse suspensions. This is an interesting open problem for future studies.

\acknowledgements
We thank Kuniyasu Saitoh for fruitful discussion.
S.T. and H.H. acknowledge the warm hospitality of the Universidad de Extremadura during their stays there.
V.G. and A.S. appreciate the warm hospitality of the Yukawa Institute for Theoretical Physics, Kyoto University, during their stays there.
The work of S.T.\ is partially supported by the Grant-in-Aid of MEXT for Scientific Research (Grant No.\ 20K14428).
The research of H.H.\ has been partially supported by the Grant-in-Aid of MEXT for Scientific Research (Grant No.\ 16H04025) and the Scholarship and ISHIZUE 2020 of Kyoto University Research Development
Program.
The research of the authors was partially supported by the YITP activities (YITP-X-19-01, YITP-T-18-03).
The research of V.G.\ and A.S.\ has been supported by the Spanish Agencia Estatal de Investigaci\'on through Grant No.\ FIS2016-76359-P and the Junta de Extremadura (Spain) through Grant No. GR18079, both partially financed by Fondo Europeo de Desarrollo Regional funds.
Numerical computation in this work was partially carried out at the Yukawa Institute Computer Facility.

\appendix

\section{Effect of the density dependence of the drag coefficient}\label{sec:R}

In the main text, we do not consider the density dependence of the drag coefficient $\zeta$.
In this Appendix, we consider such a dependence as used in Ref.\ \cite{Sangani96} to confirm whether the results are unchanged after we correct the errors in Ref.\ \cite{Hayakawa17}.
The explicit density-dependent drag coefficient is given by \cite{Hayakawa17}
\begin{equation}
	\zeta = \zeta_0 R(\varphi).
\end{equation}
Here, $\zeta_0$ in this expression corresponds to $\zeta$ in the main text.
The following form is sometimes used for the dimensionless resistance $R(\varphi)$ \cite{Sangani96, Garzo12, Hayakawa17}:
\begin{equation}
	R(\varphi)=
	\begin{cases}
	\displaystyle 1+3\sqrt{\frac{\varphi}{2}} & (\varphi\le 0.1)\\
	k_1(\varphi) - \varphi g_0(\varphi) \ln \epsilon_{\rm m} & (\varphi>0.1)
	\end{cases},
\end{equation}
with
\begin{equation}
	k_1(\varphi)
	= 1 + \frac{3}{\sqrt{2}}\varphi^{1/2} + \frac{135}{64}\varphi \ln \varphi
	+ 11.26\varphi (1-5.1\varphi+16.57\varphi^2 - 21.77\varphi^3),
\end{equation}
and $\epsilon_{\rm m}=0.01$.

We can evaluate the temperature and the shear viscosity for arbitrary shear rate case.
Following a procedure  similar to that presented in Sec.\ \ref{sec:arbitrary_shear}, we obtain a set of equations which determine the rheology for arbitrary shear rate as
\begin{subequations}
\begin{align}
	2R(\varphi)(\theta-1) + \frac{2}{3}\dot\gamma^* \theta \Pi_{xy}+\frac{1}{3}\theta \Lambda_{\alpha\alpha}^*&=0,\\
	2R(\varphi)\Pi_{xx} +\frac{4}{3}\dot\gamma^* \Pi_{xy}+\delta \Lambda_{xx}^*&=0,\\
	2R(\varphi)\Pi_{yy} -\frac{2}{3}\dot\gamma^* \Pi_{xy}+\delta \Lambda_{yy}^*&=0,\\
	2R(\varphi)\Pi_{xy} + \dot\gamma^* \left(\Pi_{yy}+1\right)+\Lambda_{xy}^*&=0.
\end{align}
\end{subequations}
Here, the expressions of $\Lambda_{\alpha\alpha}^*$, $\delta \Lambda_{xx}^*$, $\delta \Lambda_{yy}^*$, and $\Lambda_{xy}^*$ are the same as given by Eqs.\ \eqref{eq:Lii_exp-eq:Lyy_exp} in the main text.

\begin{figure}[htbp]
\includegraphics[width=\linewidth]{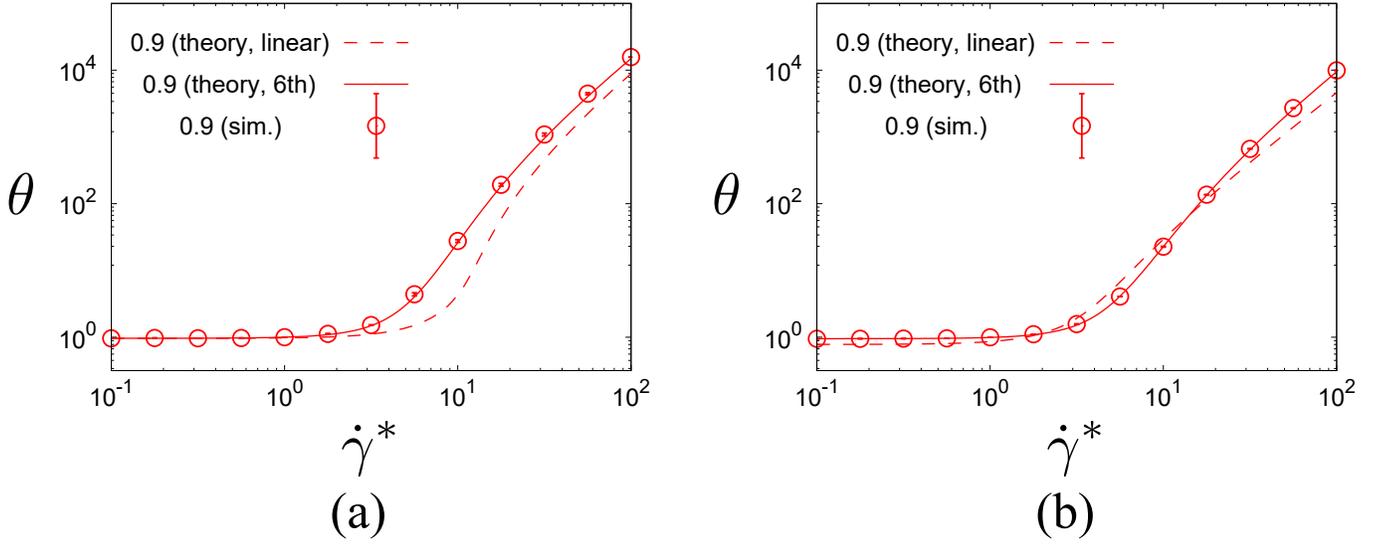}
\caption{
Plots of the dimensionless temperature $\theta$ against the shear rate for $\varphi=0.20$ and $0.30$ ($e=0.9$) when we consider the effect of the density dependence $R(\varphi)$.
The dashed and solid lines represent the results from the linear approximation and arbitrary shear case  truncated up to sixth order, respectively.
  }
\label{fig9}
\end{figure}

In Fig.\ \ref{fig9}, we present the comparison among the theory of  linear shear rate~\cite{Hayakawa17}, the theory with the sixth-order expansion, and the simulation results.
Similarly to what happens with a constant $\zeta$, we observe that the simulation results are quantitatively captured by the sixth-order expansion, 
while the linear theory works well only for $\varphi\le 0.1$.
Nevertheless, the qualitative disagreement between the linear theory~\cite{Hayakawa17} and the simulation is not large, even for $\varphi=0.3$.

\section{Derivation of formulas for arbitrary shear rate: Evaluations of $I^{(\ell)}(\widehat{\bm{\sigma}})$ and $I_\alpha^{(\ell)}(\widehat{\bm{\sigma}})$}\label{sec:derivation_I}

In this Appendix, we derive Eqs.\ \eqref{I2-I3-Jx-Jy}. 
We also note that, as in the main text, the Greek and Latin characters represent $\{x,y,z\}$ and $\{1,2,3\}$, respectively.

We start by recalling that  $I^{(\ell)}(\widehat{\bm{\sigma}})$ and $I_\alpha^{(\ell)}(\widehat{\bm{\sigma}})$ are given by Eqs.\ \eqref{eq:I_alpha-eq:I}.
Substituting Eq.\ \eqref{Maxwell} into Eq.\ \eqref{Grad}, we have
\begin{align}
	f(\bm{V}_1)f(\bm{V}_2)
	=& n^2 \left(\frac{m}{2\pi T}\right)^3 e^{-2G^2 -g^2/2}
	\left\{1+\Pi_{\alpha\beta} \left[\left(G_\alpha+g_\alpha/2\right)\left(G_\beta+g_\beta/2\right)
		 +\left(G_\alpha-g_\alpha/2\right)\left(G_\beta-g_\beta/2\right)\right]\right.\nonumber\\
	 &\left.+\Pi_{\alpha\beta}\Pi_{\gamma\delta}\left(G_\alpha+g_\alpha/2\right)\left(G_\beta+g_\beta/2\right)
		\left(G_\gamma-g_\gamma/2\right)\left(G_\delta-g_\delta/2\right)\right\},
\end{align}
where $\bm{G}\equiv (\bm{V}_1+\bm{V}_2)/2v_T$ and $\bm{g}\equiv \bm{V}_{12}/v_T$.
Then, we can write $I^{(\ell)}(\widehat{\bm{\sigma}})=n^2 v_T^\ell \tilde{I}^{(\ell)}(\widehat{\bm{\sigma}})$ and	$I_\alpha^{(\ell)}(\tilde{\bm{\sigma}})=n^2 v_T^{\ell+1} \widehat{I}_\alpha^{(\ell)}(\widehat{\bm{\sigma}})$,
where
\begin{align}
	\begin{Bmatrix} \tilde{I}^{(\ell)}(\widehat{\bm{\sigma}}) \\ \tilde{I}_\alpha^{(\ell)}(\widehat{\bm{\sigma}}) \end{Bmatrix}
	\equiv &\frac{1}{\pi^3} \int d\bm{G} \int d\bm{g} \Theta\left(\widehat{\bm{\sigma}}\cdot \bm{g} - b_T\right)
	\left(\widehat{\bm{\sigma}}\cdot \bm{g} - b_T\right)^\ell
	\begin{Bmatrix} 1 \\ g_\alpha - a_T\delta_{\alpha x} \end{Bmatrix}
	e^{-2G^2 -g^2/2}\nonumber\\
	&\times \left\{1+\Pi_{\beta\gamma} \left[\left(G_\beta+g_\beta/2\right)\left(G_\gamma+g_\gamma/2\right)
	+\left(G_\beta-g_\beta/2\right)\left(G_\gamma-g_\gamma/2\right)\right]\right.\nonumber\\
	 &\left.+\Pi_{\beta\gamma}\Pi_{\delta\mu}\left(G_\beta+g_\beta/2\right)\left(G_\gamma+g_\gamma/2\right)
	\left(G_\delta-g_\delta/2\right)\left(G_\mu-g_\mu/2\right)\right\}\nonumber\\
	=& \frac{1}{(2\pi)^{3/2}} \int d\bm{g} \Theta\left(\widehat{\bm{\sigma}}\cdot \bm{g} - b_T\right)
	\left(\widehat{\bm{\sigma}}\cdot \bm{g} - b_T\right)^\ell
	\begin{Bmatrix} 1 \\ g_\alpha - a_T\delta_{\alpha x} \end{Bmatrix}
	e^{-g^2/2} \mathcal{P}_1(\{g_\nu\}),
	\label{eq:nondim_I}
\end{align}
with
\begin{equation}
	\mathcal{P}_1(\{g_\nu\})\equiv
	1+\frac{1}{8}\Pi_{\beta\gamma}\Pi_{\beta\gamma}
	+\frac{g_\beta g_\gamma}{2}\Pi_{\beta\gamma}
	-\frac{g_\beta g_\gamma}{4}\Pi_{\beta\delta}\Pi_{\gamma\delta}
	+\frac{g_\beta g_\gamma g_\delta g_\mu}{16}\Pi_{\beta\gamma}\Pi_{\delta\mu},
\end{equation}
where we have used $\Pi_{\alpha\beta} \int d\bm{G} e^{-2G^2}G_\alpha G_\beta=0$, $\int d\bm{G}e^{-2G^2}=(\pi/2)^{3/2}$, and introduced
\begin{equation}
	a_T=\frac{a}{v_T}=\frac{\dot\gamma \sigma}{v_T} \widehat{\sigma}_y ,\quad
	b_T=\frac{b}{v_T}=\frac{\dot\gamma \sigma}{v_T} \widehat{\sigma}_x \widehat{\sigma}_y .
\end{equation}

Now, we make use of the change of basis $\{{\bm{e}}_x,{\bm{e}}_y,{\bm{e}}_z\}\to\{\bar{\bm{e}}_1,\bar{\bm{e}}_2,\bar{\bm{e}}_3\}$ with $\bar{\bm{e}}_i=U_{\alpha i}\bm{e}_\alpha$, where the  matrix $U_{\alpha i}$ is given by Eq.\ \eqref{Uai}.
Thus, $\bm{g}=g_{\alpha}\bm{e}_\alpha=\bar{g}_{i}\bar{\bm{e}}_i$ with $g_{\alpha}=U_{\alpha i}\bar{g}_{i}$ and $\bar{g}_{i}=U_{\alpha i}g_{\alpha}$.
Using these variables, we can rewrite Eq.\ \eqref{eq:nondim_I} as
\begin{equation}
	\begin{Bmatrix} \tilde{I}^{(\ell)}(\tilde{\bm{\sigma}}) \\ \tilde{I}_\alpha^{(\ell)}(\widehat{\bm{\sigma}}) \end{Bmatrix}
	= \frac{1}{(2\pi)^{3/2}} \int d{\bm{g}} \Theta\left(\bar{g}_3- b_T\right)
	\left(\bar{g}_3 - b_T\right)^\ell
	\begin{Bmatrix} 1 \\ U_{\alpha i}\bar{g}_i - a_T\delta_{\alpha x} \end{Bmatrix}
	e^{-\bar{g}^2/2} \mathcal{P}_2(\{\bar{g}_m\};\widehat{\bm{\sigma}}),
	\label{eq:I_alpha_ell}
\end{equation}
with
\begin{align}
\label{PP2}
	\mathcal{P}_2(\{\bar{g}_m\};\widehat{\bm{\sigma}})
	&=\mathcal{P}_1(\{U_{\nu m}\bar{g}_m\})\nonumber\\
	&=
	1+\frac{1}{8}\Pi_{\beta\gamma}\Pi_{\beta\gamma}
	+\frac{1}{2}\bar{g}_i \bar{g}_j U_{\beta i} U_{\gamma j} \Pi_{\beta\gamma}
	-\frac{1}{4}\bar{g}_i \bar{g}_j U_{\beta i} U_{\gamma j} \Pi_{\beta\delta}\Pi_{\gamma\delta}
	+\frac{1}{16}\bar{g}_i \bar{g}_j\bar{g}_k \bar{g}_l U_{\beta i} U_{\gamma j} U_{\delta k} U_{\mu l}\Pi_{\beta\gamma}\Pi_{\delta\mu}.
\end{align}

Analogously to what was done in Eqs.\ \eqref{Ja} and \eqref{eq:relation_I2_I3_J}, it is convenient to expand $U_{\alpha i}\bar{g}_i$ and decompose $\tilde{I}_\alpha^{(\ell)}(\widehat{\bm{\sigma}})$ as
\begin{align}
\label{C7}
\tilde{I}_\alpha^{(\ell)}(\widehat{\bm{\sigma}})=U_{\alpha 1}\tilde{\bar{I}}_1^{(\ell)}(\widehat{\bm{\sigma}})+U_{\alpha 2}\tilde{\bar{I}}_2^{(\ell)}(\widehat{\bm{\sigma}})+ \widehat{\sigma}_\alpha \tilde{I}^{(\ell+1)}\left(\widehat{\bm{\sigma}}\right)+a_T\left(\widehat{\sigma}_\alpha\widehat{\sigma}_x- \delta_{\alpha x}\right)\tilde{I}^{(\ell)}\left(\widehat{\bm{\sigma}}\right),
\end{align}
where
\begin{equation}
\tilde{\bar{I}}_i^{(\ell)}(\widehat{\bm{\sigma}}) = \frac{1}{(2\pi)^{3/2}} \int d{\bm{g}} \Theta\left(\bar{g}_3- b_T\right)
	\left(\bar{g}_3 - b_T\right)^\ell
\bar{g}_i
	e^{-\bar{g}^2/2} \mathcal{P}_2(\{\bar{g}_m\};\widehat{\bm{\sigma}}),\quad (i=1,2).
	\label{eq:barI_alpha_ell}
\end{equation}
Integrating over $\bar{g}_1$ and $\bar{g}_2$, we have
\begin{equation}
	\begin{Bmatrix} \tilde{I}^{(\ell)}(\tilde{\bm{\sigma}}) \\ \tilde{\bar{I}}_i^{(\ell)}(\widehat{\bm{\sigma}}) \end{Bmatrix}
	= \frac{1}{\sqrt{2\pi}} \int_{b_T}^\infty d{\bar{g}_3}
	\left(\bar{g}_3 - b_T\right)^\ell e^{-\bar{g}_3^2/2}
	\begin{Bmatrix} \mathcal{P}_3(\bar{g}_3;\widehat{\bm{\sigma}}) \\ \mathcal{P}_{3,i}(\bar{g}_3;\widehat{\bm{\sigma}})  \end{Bmatrix}
	 ,
	\label{eq:def_I}
\end{equation}
where
\begin{equation}
\label{PP3}
	\begin{Bmatrix}
\mathcal{P}_{3}(\bar{g}_3;\widehat{\bm{\sigma}})\\
\mathcal{P}_{3,i}(\bar{g}_3;\widehat{\bm{\sigma}})\end{Bmatrix}
	\equiv \frac{1}{2\pi} \int_{-\infty}^\infty d\bar{g}_1 \int_{-\infty}^\infty d\bar{g}_2
	e^{-\bar{g}_1^2/2 - \bar{g}_2^2/2} \begin{Bmatrix}
1\\
\bar{g}_i \end{Bmatrix}
\mathcal{P}_2(\{\bar{g}_m\};\widehat{\bm{\sigma}}).
\end{equation}
Inserting Eq.\ \eqref{PP2} into Eq.\ \eqref{PP3} one gets
\begin{subequations}
\begin{align}
\mathcal{P}_{3}(\bar{g}_3;\widehat{\bm{\sigma}})=&\mathcal{P}_{3}^{(0)}(\bar{g}_3;\widehat{\bm{\sigma}})+
\mathcal{P}_{3}^{(1)}(\bar{g}_3;\widehat{\bm{\sigma}})+\mathcal{P}_{3}^{(2)}(\bar{g}_3;\widehat{\bm{\sigma}})+\mathcal{P}_{3}^{(3)}(\bar{g}_3;\widehat{\bm{\sigma}})
\label{C11a}\\
\mathcal{P}_{3,i}(\bar{g}_3;\widehat{\bm{\sigma}})=&
\mathcal{P}_{3,i}^{(1)}(\bar{g}_3;\widehat{\bm{\sigma}})+\mathcal{P}_{3,i}^{(2)}(\bar{g}_3;\widehat{\bm{\sigma}})+\mathcal{P}_{3,i}^{(3)}(\bar{g}_3;\widehat{\bm{\sigma}})
,
\end{align}
\end{subequations}
where
\begin{subequations}
\label{C12}
\begin{align}
\mathcal{P}_3^{(0)}(\bar{g}_3;\widehat{\bm{\sigma}})
	&\equiv \frac{1}{2\pi} \int_{-\infty}^\infty d\bar{g}_1 \int_{-\infty}^\infty d\bar{g}_2
	e^{-\bar{g}_1^2/2 - \bar{g}_2^2/2}
	\left(1+\frac{1}{8}\Pi_{\beta\gamma}\Pi_{\beta\gamma}\right),\\
	\begin{Bmatrix}
\mathcal{P}_{3}^{(1)}(\bar{g}_3;\widehat{\bm{\sigma}})\\
\mathcal{P}_{3,1}^{(1)}(\bar{g}_3;\widehat{\bm{\sigma}})\\
\mathcal{P}_{3,2}^{(1)}(\bar{g}_3;\widehat{\bm{\sigma}})
\end{Bmatrix}
	&\equiv \frac{1}{4\pi} \int_{-\infty}^\infty d\bar{g}_1 \int_{-\infty}^\infty d\bar{g}_2
	e^{-\bar{g}_1^2/2 - \bar{g}_2^2/2}
\begin{Bmatrix}
1\\
\bar{g}_1\\
\bar{g}_2
\end{Bmatrix}
\bar{g}_i \bar{g}_j U_{\beta i} U_{\gamma j} \Pi_{\beta\gamma},\\
	\begin{Bmatrix}
\mathcal{P}_{3}^{(2)}(\bar{g}_3;\widehat{\bm{\sigma}})\\
\mathcal{P}_{3,1}^{(2)}(\bar{g}_3;\widehat{\bm{\sigma}})\\
\mathcal{P}_{3,2}^{(2)}(\bar{g}_3;\widehat{\bm{\sigma}})
\end{Bmatrix}
	&\equiv -\frac{1}{8\pi} \int_{-\infty}^\infty d\bar{g}_1 \int_{-\infty}^\infty d\bar{g}_2
	e^{-\bar{g}_1^2/2 - \bar{g}_2^2/2}\begin{Bmatrix}
1\\
\bar{g}_1\\
\bar{g}_2
\end{Bmatrix}
	\bar{g}_i \bar{g}_j U_{\beta i} U_{\gamma j} \Pi_{\beta\delta}\Pi_{\gamma\delta},\\
	\begin{Bmatrix}
\mathcal{P}_{3}^{(3)}(\bar{g}_3;\widehat{\bm{\sigma}})\\
\mathcal{P}_{3,1}^{(3)}(\bar{g}_3;\widehat{\bm{\sigma}})
\\
\mathcal{P}_{3,2}^{(3)}(\bar{g}_3;\widehat{\bm{\sigma}})
\end{Bmatrix}
	&\equiv \frac{1}{32\pi} \int_{-\infty}^\infty d\bar{g}_1 \int_{-\infty}^\infty d\bar{g}_2
	e^{-\bar{g}_1^2/2 - \bar{g}_2^2/2}
\begin{Bmatrix}
1\\
\bar{g}_1\\
\bar{g}_2
\end{Bmatrix}
	\bar{g}_i \bar{g}_j\bar{g}_k \bar{g}_l U_{\beta i} U_{\gamma j} U_{\delta k} U_{\mu l}\Pi_{\beta\gamma}\Pi_{\delta\mu}.
\end{align}
\end{subequations}
Now we proceed to the evaluation of $\mathcal{P}_{3,1}(\bar{g}_3;\widehat{\bm{\sigma}})$, $\mathcal{P}_{3,2}(\bar{g}_3;\widehat{\bm{\sigma}})$, and $\mathcal{P}_{3}(\bar{g}_3;\widehat{\bm{\sigma}})$.

\subsection{Evaluation of $\mathcal{P}_{3,1}(\bar{g}_3;\widehat{\bm{\sigma}})$}

By using symmetry properties, the first and second contributions to $\mathcal{P}_{3,1}(\bar{g}_3;\widehat{\bm{\sigma}})$ are calculated as
\begin{subequations}
\label{eq:P31_1,2}
\begin{align}
	\mathcal{P}_{3,1}^{(1)}(\bar{g}_3;\widehat{\bm{\sigma}})
	&= \frac{1}{4\pi} \int_{-\infty}^\infty d\bar{g}_1 \int_{-\infty}^\infty d\bar{g}_2
	e^{-\bar{g}_1^2/2 - \bar{g}_2^2/2}
	 2\bar{g}_1^2 \bar{g}_3 U_{\beta 1} U_{\gamma 3} \Pi_{\beta\gamma}\nonumber\\
	&= \bar{g}_3 U_{\beta 1} U_{\gamma 3} \Pi_{\beta\gamma}\nonumber\\
		&= \frac{\bar{g}_3}{\sqrt{\widehat{\sigma}_x^2+\widehat{\sigma}_y^2}}
	\left[\widehat{\sigma}_x \widehat{\sigma}_y \left(\Pi_{xx} -\Pi_{yy}\right) -\left(\widehat{\sigma}_x^2-\widehat{\sigma}_y^2\right) \Pi_{xy}\right],\\
	\mathcal{P}_{3,1}^{(2)}(\bar{g}_3;\widehat{\bm{\sigma}})
	&= -\frac{1}{8\pi} \int_{-\infty}^\infty d\bar{g}_1 \int_{-\infty}^\infty d\bar{g}_2
	e^{-\bar{g}_1^2/2 - \bar{g}_2^2/2}
	2\bar{g}_1^2 \bar{g}_3 U_{\beta 1} U_{\gamma 3} \Pi_{\beta\delta}\Pi_{\gamma\delta}\nonumber\\
	&= -\frac{1}{2}\bar{g}_3 U_{\beta 1} U_{\gamma 3} \Pi_{\beta\delta}\Pi_{\gamma\delta}\nonumber\\
	&= -\frac{1}{2}\frac{\bar{g}_3}{\sqrt{\widehat{\sigma}_x^2+\widehat{\sigma}_y^2}}
	\left[\widehat{\sigma}_x \widehat{\sigma}_y \left(\Pi_{xx} -\Pi_{yy}\right) -\left(\widehat{\sigma}_x^2-\widehat{\sigma}_y^2\right) \Pi_{xy}\right]
	\left(\Pi_{xx} + \Pi_{yy}\right).
\end{align}
\end{subequations}
In the third equalities of Eqs.\ \eqref{eq:P31_1,2} we have made use of the explicit form of the tensor $U_{\alpha i}$ [see Eq.\ \eqref{Uai}] and of the relation $\Pi_{zz}=-\Pi_{xx}-\Pi_{yy}$. This will also be done in the remainder of this Appendix.

By symmetry, the third contribution, $\mathcal{P}_{3,1}^{(3)}(\bar{g}_3;\widehat{\bm{\sigma}})$, is made of four terms, namely
\begin{equation}
\label{eq:P31_2}
\mathcal{P}_{3,1}^{(3)}(\bar{g}_3;\widehat{\bm{\sigma}})=\mathcal{P}_{3,1}^{(3,1)}(\bar{g}_3;\widehat{\bm{\sigma}})
	+\mathcal{P}_{3,1}^{(3,2)}(\bar{g}_3;\widehat{\bm{\sigma}})
	+\mathcal{P}_{3,1}^{(3,3)}(\bar{g}_3;\widehat{\bm{\sigma}})
	+\mathcal{P}_{3,1}^{(3,4)}(\bar{g}_3;\widehat{\bm{\sigma}}),
\end{equation}
where
\begin{subequations}
\label{eq:P31_3i}
\begin{align}
	\mathcal{P}_{3,1}^{(3,1)}(\bar{g}_3;\widehat{\bm{\sigma}})
	&\equiv \frac{1}{32\pi} \int_{-\infty}^\infty d\bar{g}_1 \int_{-\infty}^\infty d\bar{g}_2
	e^{-\bar{g}_1^2/2 - \bar{g}_2^2/2}4\bar{g}_1^4\bar{g}_3 U_{\beta 1}U_{\gamma 1} U_{\delta 1} U_{\mu 3}\Pi_{\beta\gamma} \Pi_{\delta\mu}\nonumber\\
    &=\frac{3}{4}\bar{g}_3 U_{\beta 1}U_{\gamma 1} U_{\delta 1} U_{\mu 3}\Pi_{\beta\gamma} \Pi_{\delta\mu}\nonumber\\
    &= \frac{3}{4}\frac{\bar{g}_3}{\left(\widehat{\sigma}_x^2+\widehat{\sigma}_y^2\right)^{3/2}}
	\left[\widehat{\sigma}_x \widehat{\sigma}_y \left(\Pi_{xx} -\Pi_{yy}\right) - \left(\widehat{\sigma}_x^2 - \widehat{\sigma}_y^2\right)\Pi_{xy}\right]
	\left[-\widehat{\sigma}_\beta\widehat{\sigma}_\gamma\Pi_{\beta\gamma}+\left(1-2\widehat{\sigma}_z^2\right)\left(\Pi_{xx}+ \Pi_{yy}\right)\right],
	\label{eq:P31_31}\\
   \mathcal{P}_{3,1}^{(3,2)}(\bar{g}_3;\widehat{\bm{\sigma}})
	&\equiv \frac{1}{32\pi} \int_{-\infty}^\infty d\bar{g}_1 \int_{-\infty}^\infty d\bar{g}_2
	e^{-\bar{g}_1^2/2 - \bar{g}_2^2/2}4\bar{g}_1^2\bar{g}_3^3 U_{\beta 1}U_{\gamma 3} U_{\delta 3} U_{\mu 3}\Pi_{\beta\gamma} \Pi_{\delta\mu}\nonumber\\
    &=\frac{1}{4}\bar{g}_3^3 U_{\beta 1}U_{\gamma 3} U_{\delta 3} U_{\mu 3}\Pi_{\beta\gamma} \Pi_{\delta\mu}\nonumber\\
    &= \frac{1}{4}\frac{\bar{g}_3^3}{\sqrt{\widehat{\sigma}_x^2+\widehat{\sigma}_y^2}}
	\left[\widehat{\sigma}_x \widehat{\sigma}_y \left(\Pi_{xx} -\Pi_{yy}\right) - \left(\widehat{\sigma}_x^2 - \widehat{\sigma}_y^2\right)\Pi_{xy}\right]
	\widehat{\sigma}_\beta\widehat{\sigma}_\gamma\Pi_{\beta\gamma},
	\label{eq:P31_32}\\
    \mathcal{P}_{3,1}^{(3,3)}(\bar{g}_3;\widehat{\bm{\sigma}})
	&\equiv \frac{1}{32\pi} \int_{-\infty}^\infty d\bar{g}_1 \int_{-\infty}^\infty d\bar{g}_2
	e^{-\bar{g}_1^2/2 - \bar{g}_2^2/2}8\bar{g}_1^2\bar{g}_2^2\bar{g}_3 U_{\beta 1}U_{\gamma 2} U_{\delta 2} U_{\mu 3}\Pi_{\beta\gamma} \Pi_{\delta\mu}\nonumber\\
    &=\frac{1}{2}\bar{g}_3U_{\beta 1}U_{\gamma 2} U_{\delta 2} U_{\mu 3}\Pi_{\beta\gamma} \Pi_{\delta\mu}\nonumber\\
    &= \frac{1}{2} \frac{\bar{g}_3\widehat{\sigma}_z^2}{\left(\widehat{\sigma}_x^2+\widehat{\sigma}_y^2\right)^{3/2}}
	\left[\widehat{\sigma}_x \widehat{\sigma}_y \left(\Pi_{xx} -\Pi_{yy}\right) - \left(\widehat{\sigma}_x^2 - \widehat{\sigma}_y^2\right)\Pi_{xy}\right]
	\left( \widehat{\sigma}_\beta\widehat{\sigma}_\gamma\Pi_{\beta\gamma}+\Pi_{xx} + \Pi_{yy}\right),
	\label{eq:P31_33}\\
    \mathcal{P}_{3,1}^{(3,4)}(\bar{g}_3;\widehat{\bm{\sigma}})
	&\equiv \frac{1}{32\pi} \int_{-\infty}^\infty d\bar{g}_1 \int_{-\infty}^\infty d\bar{g}_2
	e^{-\bar{g}_1^2/2 - \bar{g}_2^2/2}4\bar{g}_1^2\bar{g}_2^2 \bar{g}_3 U_{\beta 1}U_{\gamma 3} U_{\delta 2} U_{\mu 2}\Pi_{\beta\gamma} \Pi_{\delta\mu}\nonumber\\
    &=\frac{1}{4}\bar{g}_3U_{\beta 1}U_{\gamma 3} U_{\delta 2} U_{\mu 2}\Pi_{\beta\gamma} \Pi_{\delta\mu}\nonumber\\
    &= \frac{1}{4} \frac{\bar{g}_3}{\left(\widehat{\sigma}_x^2+\widehat{\sigma}_y^2\right)^{3/2}}
	\left[\widehat{\sigma}_x \widehat{\sigma}_y \left(\Pi_{xx} -\Pi_{yy}\right) - \left(\widehat{\sigma}_x^2 - \widehat{\sigma}_y^2\right)\Pi_{xy}\right]
	\left[\widehat{\sigma}_z^2\widehat{\sigma}_\beta\widehat{\sigma}_\gamma\Pi_{\beta\gamma}
	-\left(1-2\widehat{\sigma}_z^2\right) \left(\Pi_{xx}+\Pi_{yy}\right)\right].
\label{eq:P31_34}
\end{align}
\end{subequations}
Combining these four terms, the result is
\begin{equation}
	\mathcal{P}_{3,1}^{(3)}(\bar{g}_3;\widehat{\bm{\sigma}})
	= \frac{1}{4}\frac{\bar{g}_3}{\sqrt{\widehat{\sigma}_x^2+\widehat{\sigma}_y^2}}
	\left[\widehat{\sigma}_x \widehat{\sigma}_y \left(\Pi_{xx} -\Pi_{yy}\right) - \left(\widehat{\sigma}_x^2 - \widehat{\sigma}_y^2\right)\Pi_{xy}\right]
	\left[\left(\bar{g}_3^2-3\right)\widehat{\sigma}_\beta\widehat{\sigma}_\gamma\Pi_{\beta\gamma}
	+2\left(\Pi_{xx}+\Pi_{yy}\right)\right].
	\label{eq:P31_3}
\end{equation}

Finally, summing up Eqs.\ \eqref{eq:P31_1,2} and \eqref{eq:P31_3}, we obtain the expression of $\mathcal{P}_{3,1}(\bar{g}_3;\widehat{\bm{\sigma}})$ as
\begin{equation}
	\mathcal{P}_{3,1}(\bar{g}_3;\widehat{\bm{\sigma}})
	=\frac{\bar{g}_3}{\sqrt{\widehat{\sigma}_x^2+\widehat{\sigma}_y^2}} 	
	\left(1+\frac{\bar{g}_3^2-3}{4}\widehat{\sigma}_\beta\widehat{\sigma}_\gamma\Pi_{\beta\gamma}\right)
	\left[\widehat{\sigma}_x \widehat{\sigma}_y \left(\Pi_{xx} -\Pi_{yy}\right) -\left(\widehat{\sigma}_x^2-\widehat{\sigma}_y^2\right)\Pi_{xy}\right].
	\label{eq:P31}
\end{equation}

\subsection{Evaluation of $\mathcal{P}_{3,2}(\bar{g}_3;\widehat{\bm{\sigma}})$}

Using the same procedure, we can derive the expression of $\mathcal{P}_{3,2}(\bar{g}_3;\widehat{\bm{\sigma}})$. First, note that the first and second equalities in each one of Eqs.\ \eqref{eq:P31_1,2} apply to $\mathcal{P}_{3,2}^{(1)}(\bar{g}_3;\widehat{\bm{\sigma}})$ and $\mathcal{P}_{3,2}^{(2)}(\bar{g}_3;\widehat{\bm{\sigma}})$ just by the exchange of indices $1\leftrightarrow 2$, so that
\begin{subequations}
\label{eq:P32_1,2}
\begin{align}
	\mathcal{P}_{3,2}^{(1)}(\bar{g}_3;\widehat{\bm{\sigma}})
		&= \bar{g}_3 U_{\beta 2} U_{\gamma 3} \Pi_{\beta\gamma}\nonumber\\
		&=  \frac{\bar{g}_3 \widehat{\sigma}_z}{\sqrt{\widehat{\sigma}_x^2+\widehat{\sigma}_y^2}}
	\left(\widehat{\sigma}_\beta \widehat{\sigma}_\gamma \Pi_{\beta\gamma} +\Pi_{xx} + \Pi_{yy}\right),\\
	\mathcal{P}_{3,2}^{(2)}(\bar{g}_3;\widehat{\bm{\sigma}})
		&= -\frac{1}{2}\bar{g}_3 U_{\beta 2} U_{\gamma 3} \Pi_{\beta\delta}\Pi_{\gamma\delta}\nonumber\\
	&=  -\frac{1}{2} \frac{\bar{g}_3\widehat{\sigma}_z}{\sqrt{\widehat{\sigma}_x^2 + \widehat{\sigma}_y^2}}
	\left[\widehat{\sigma}_x^2 \left(\Pi_{xx}^2 - \Pi_{zz}^2\right) + \widehat{\sigma}_y^2 \left(\Pi_{yy}^2 - \Pi_{zz}^2\right)
	+ \left(\widehat{\sigma}_x^2 +\widehat{\sigma}_y^2\right) \Pi_{xy}^2 + 2\widehat{\sigma}_x \widehat{\sigma}_y \Pi_{xy}\left(\Pi_{xx}+\Pi_{yy}\right)\right].
\end{align}
\end{subequations}

Analogously to Eqs.\ \eqref{eq:P31_2} and \eqref{eq:P31_3i}, $\mathcal{P}_{3,2}^{(3)}(\bar{g}_3;\widehat{\bm{\sigma}})$ is the sum of four terms:
\begin{subequations}
\label{eq:P32_3i}
\begin{align}
\label{eq:P32_31}
	\mathcal{P}_{3,2}^{(3,1)}(\bar{g}_3;\widehat{\bm{\sigma}})
	    &=\frac{3}{4}\bar{g}_3 U_{\beta 2}U_{\gamma 2} U_{\delta 2} U_{\mu 3}\Pi_{\beta\gamma} \Pi_{\delta\mu}\nonumber\\
&=\frac{3}{4}\frac{\bar{g}_3 \widehat{\sigma}_z}{\left(\widehat{\sigma}_x^2+\widehat{\sigma}_y^2\right)^{3/2}}
	\left(\widehat{\sigma}_\beta \widehat{\sigma}_\gamma \Pi_{\beta\gamma} +\Pi_{xx} + \Pi_{yy}\right)
	\left[\widehat{\sigma}_z^2 \widehat{\sigma}_\delta \widehat{\sigma}_\mu \Pi_{\delta \mu}
	-\left(1-2\widehat{\sigma}_z^2\right) \left(\Pi_{xx}+\Pi_{yy}\right)\right],\\
\label{eq:P32_32}
   \mathcal{P}_{3,2}^{(3,2)}(\bar{g}_3;\widehat{\bm{\sigma}})
	&=\frac{1}{4}\bar{g}_3^3 U_{\beta 2}U_{\gamma 3} U_{\delta 3} U_{\mu 3}\Pi_{\beta\gamma} \Pi_{\delta\mu}\nonumber\\
&=\frac{1}{4}\frac{\bar{g}_3^3 \widehat{\sigma}_z}{\sqrt{\widehat{\sigma}_x^2+\widehat{\sigma}_y^2}}
	\left(\widehat{\sigma}_\beta \widehat{\sigma}_\gamma \Pi_{\beta\gamma} +\Pi_{xx} + \Pi_{yy}\right)
	\widehat{\sigma}_\delta \widehat{\sigma}_\mu \Pi_{\delta\mu},\\
    \mathcal{P}_{3,2}^{(3,3)}(\bar{g}_3;\widehat{\bm{\sigma}})
	&=\frac{1}{2}\bar{g}_3U_{\beta 2}U_{\gamma 1} U_{\delta 1} U_{\mu 3}\Pi_{\beta\gamma} \Pi_{\delta\mu}\nonumber\\
    &= \frac{1}{2} \frac{\bar{g}_3\widehat{\sigma}_z}{\left(\widehat{\sigma}_x^2+\widehat{\sigma}_y^2\right)^{3/2}}
	\left[\widehat{\sigma}_x \widehat{\sigma}_y \left(\Pi_{xx} -\Pi_{yy}\right) - \left(\widehat{\sigma}_x^2 - \widehat{\sigma}_y^2\right)\Pi_{xy}\right]^2,
	\label{eq:P32_33}\\
    \mathcal{P}_{3,2}^{(3,4)}(\bar{g}_3;\widehat{\bm{\sigma}})
    &=\frac{1}{4}\bar{g}_3U_{\beta 2}U_{\gamma 3} U_{\delta 1} U_{\mu 1}\Pi_{\beta\gamma} \Pi_{\delta\mu}\nonumber\\
    &= \frac{1}{4}\frac{\bar{g}_3 \widehat{\sigma}_z}{\left(\widehat{\sigma}_x^2+\widehat{\sigma}_y^2\right)^{3/2}}
	\left(\widehat{\sigma}_\beta \widehat{\sigma}_\gamma \Pi_{\beta \gamma} + \Pi_{xx} + \Pi_{yy}\right)
	\left[-\widehat{\sigma}_\delta \widehat{\sigma}_\mu \Pi_{\delta\mu} +\left(1-2\widehat{\sigma}_z^2\right)\left(\Pi_{xx}+\Pi_{yy}\right)\right].
\label{eq:P32_34}
\end{align}
\end{subequations}

{}From Eqs.\ \eqref{eq:P32_1,2} and \eqref{eq:P32_3i}, and after some algebra, we obtain
\begin{subequations}
\label{eq:P32_sum1-P32_sum2}
\begin{align}
	 \mathcal{P}_{3,2}^{(1)}(\bar{g}_3;\widehat{\bm{\sigma}})+\mathcal{P}_{3,2}^{(3,2)}(\bar{g}_3;\widehat{\bm{\sigma}})
	=& \frac{\bar{g}_3 \widehat{\sigma}_z}{\sqrt{\widehat{\sigma}_x^2+\widehat{\sigma}_y^2}}
	\left(\widehat{\sigma}_\beta \widehat{\sigma}_\gamma \Pi_{\beta\gamma} +\Pi_{xx}+ \Pi_{yy}\right)
	\left(1+\frac{\bar{g}_3^2}{4}\widehat{\sigma}_\delta \widehat{\sigma}_\mu \Pi_{\delta\mu}\right),
	\label{eq:P32_sum1}\\
 \mathcal{P}_{3,2}^{(2)}(\bar{g}_3;\widehat{\bm{\sigma}})+\mathcal{P}_{3,2}^{(3,3)}(\bar{g}_3;\widehat{\bm{\sigma}})
	=& -\frac{1}{2}\frac{\bar{g}_3 \widehat{\sigma}_z}{(\widehat{\sigma}_x^2+\widehat{\sigma}_y^2)^{3/2}}
	\left(\widehat{\sigma}_\beta \widehat{\sigma}_\gamma \Pi_{\beta\gamma} +\Pi_{xx} + \Pi_{yy}\right)
	\left[\widehat{\sigma}_\delta \widehat{\sigma}_\mu \Pi_{\delta \mu} - \left(1-2\widehat{\sigma}_z^2\right)\left(\Pi_{xx}+\Pi_{yy}\right)\right].
	\label{eq:P32_sum3}\\
	 \mathcal{P}_{3,2}^{(3,1)}(\bar{g}_3;\widehat{\bm{\sigma}})+\mathcal{P}_{3,2}^{(3,4)}(\bar{g}_3;\widehat{\bm{\sigma}})
		=& -\frac{1}{4}\frac{\bar{g}_3 \widehat{\sigma}_z}{(\widehat{\sigma}_x^2+\widehat{\sigma}_y^2)^{3/2}}
	\left(\widehat{\sigma}_\beta \widehat{\sigma}_\gamma \Pi_{\beta\gamma}+\Pi_{xx} + \Pi_{yy}\right)
	\left[\left(1-3\widehat{\sigma}_z^2\right)\widehat{\sigma}_\delta \widehat{\sigma}_\mu \Pi_{\delta\mu}\right.\nonumber\\
	&\left.+2\left(1-2\widehat{\sigma}_z^2\right)\left(\Pi_{xx}+\Pi_{yy}\right)\right].
	\label{eq:P32_sum2}
\end{align}
\end{subequations}
Summing up all the terms in Eqs.\ \eqref{eq:P32_sum1-P32_sum2}, we finally find
\begin{equation}
	\mathcal{P}_{3,2}(\bar{g}_3;\widehat{\bm{\sigma}})
	=
	\frac{\bar{g}_3 \widehat{\sigma}_z}{\sqrt{\widehat{\sigma}_x^2+\widehat{\sigma}_y^2}}
	\left(\widehat{\sigma}_\beta \widehat{\sigma}_\gamma \Pi_{\beta\gamma} +\Pi_{xx} + \Pi_{yy}\right)
	\left(1+\frac{\bar{g}_3^2-3}{4}\widehat{\sigma}_\delta \widehat{\sigma}_\mu \Pi_{\delta\mu} \right).
	\label{eq:P32}
\end{equation}

\subsection{Evaluation of $\mathcal{P}_{3}(\bar{g}_3;\widehat{\bm{\sigma}})$}

The four contributions to $\mathcal{P}_{3}(\bar{g}_3;\widehat{\bm{\sigma}})$ are given by Eqs.\ \eqref{C11a} and \eqref{C12}. The contribution  $\mathcal{P}_3^{(0)}(\bar{g}_3;\widehat{\bm{\sigma}})$ is straightforward:
\begin{equation}
	\mathcal{P}_3^{(0)}(\bar{g}_3;\widehat{\bm{\sigma}})
	= 1+\frac{1}{8}\Pi_{\beta\gamma}\Pi_{\beta\gamma}.
	\label{eq:P3_1}
\end{equation}
Next, we decompose
\begin{equation}
	\mathcal{P}_3^{(1)}(\bar{g}_3;\widehat{\bm{\sigma}})
	= \mathcal{P}_3^{(1,1)}(\bar{g}_3;\widehat{\bm{\sigma}})
	+\mathcal{P}_3^{(1,2)}(\bar{g}_3;\widehat{\bm{\sigma}})
	+\mathcal{P}_3^{(1,3)}(\bar{g}_3;\widehat{\bm{\sigma}}),
\end{equation}
where
\begin{subequations}
\begin{align}
	\mathcal{P}_3^{(1,1)}(\bar{g}_3;\widehat{\bm{\sigma}})
	\equiv&
\frac{1}{4\pi} \int_{-\infty}^\infty d\bar{g}_1 \int_{-\infty}^\infty d\bar{g}_2
	e^{-\bar{g}_1^2/2 - \bar{g}_2^2/2}
\bar{g}_1^2  U_{\beta 1} U_{\gamma 1} \Pi_{\beta\gamma}\nonumber\\
=&\frac{1}{2} U_{\beta 1} U_{\gamma 1}\Pi_{\beta\gamma}
	= \frac{1}{2} \frac{1}{\widehat{\sigma}_x^2+\widehat{\sigma}_y^2} \left[-\widehat{\sigma}_\beta \widehat{\sigma}_\gamma \Pi_{\beta\gamma} +\left(1-2\widehat{\sigma}_z^2\right)\left(\Pi_{xx}+\Pi_{yy}\right)\right],
	\label{eq:P3_21}\\
\mathcal{P}_3^{(1,2)}(\bar{g}_3;\widehat{\bm{\sigma}})
	\equiv&
\frac{1}{4\pi} \int_{-\infty}^\infty d\bar{g}_1 \int_{-\infty}^\infty d\bar{g}_2
	e^{-\bar{g}_1^2/2 - \bar{g}_2^2/2}
\bar{g}_2^2  U_{\beta 2} U_{\gamma 2} \Pi_{\beta\gamma}\nonumber\\
=&\frac{1}{2} U_{\beta 2} U_{\gamma 2}\Pi_{\beta\gamma}
	= \frac{1}{2} \frac{1}{\widehat{\sigma}_x^2 + \widehat{\sigma}_y^2}
	\left[\widehat{\sigma}_z^2 \widehat{\sigma}_\beta \widehat{\sigma}_\gamma \Pi_{\beta \gamma}
	-\left(1-2\widehat{\sigma}_z^2\right) \left(\Pi_{xx}+\Pi_{yy}\right)\right],
	\label{eq:P3_22}\\
\mathcal{P}_3^{(1,3)}(\bar{g}_3;\widehat{\bm{\sigma}})
	\equiv&
\frac{1}{4\pi} \int_{-\infty}^\infty d\bar{g}_1 \int_{-\infty}^\infty d\bar{g}_2
	e^{-\bar{g}_1^2/2 - \bar{g}_2^2/2}
\bar{g}_3^2  U_{\beta 3} U_{\gamma 3} \Pi_{\beta\gamma}\nonumber\\
=&\frac{1}{2} \bar{g}_3^2 U_{\beta 3} U_{\gamma 3} \Pi_{\beta\gamma}
	= \frac{1}{2} \bar{g}_3^2 \widehat{\sigma}_\beta \widehat{\sigma}_\gamma \Pi_{\beta \gamma}.
	\label{eq:P3_23}
\end{align}
\end{subequations}
Thus,
\begin{equation}
	\mathcal{P}_3^{(1)}(\bar{g}_3;\widehat{\bm{\sigma}})
	= \frac{\bar{g}_3^2-1}{2}\widehat{\sigma}_\beta \widehat{\sigma}_\gamma \Pi_{\beta \gamma}.
	\label{eq:P3_2}
\end{equation}

Analogously,
\begin{equation}
	\mathcal{P}_3^{(2)}(\bar{g}_3;\widehat{\bm{\sigma}})
	= \mathcal{P}_3^{(2,1)}(\bar{g}_3;\widehat{\bm{\sigma}})
	+\mathcal{P}_3^{(2,2)}(\bar{g}_3;\widehat{\bm{\sigma}})
	+\mathcal{P}_3^{(2,3)}(\bar{g}_3;\widehat{\bm{\sigma}}),
\end{equation}
where
\begin{subequations}
\begin{align}
	\mathcal{P}_3^{(2,1)}(\bar{g}_3;\widehat{\bm{\sigma}})
	\equiv&-
\frac{1}{8\pi} \int_{-\infty}^\infty d\bar{g}_1 \int_{-\infty}^\infty d\bar{g}_2
	e^{-\bar{g}_1^2/2 - \bar{g}_2^2/2}
\bar{g}_1^2  U_{\beta 1} U_{\gamma 1} \Pi_{\beta\delta}\Pi_{\gamma\delta}\nonumber\\
=&-\frac{1}{4} U_{\beta 1} U_{\gamma 1}\Pi_{\beta\delta}\Pi_{\gamma\delta}\nonumber\\
=& -\frac{1}{4} \frac{1}{\widehat{\sigma}_x^2 + \widehat{\sigma}_y^2}\left[
	\widehat{\sigma}_y^2 \Pi_{xx}^2 + \widehat{\sigma}_x^2 \Pi_{yy}^2
	+ \left(\widehat{\sigma}_x^2+ \widehat{\sigma}_y^2\right)\Pi_{xy}^2 - 2\widehat{\sigma}_x \widehat{\sigma}_y \Pi_{xy}\left(\Pi_{xx} + \Pi_{yy}\right)
	\right],\\
\mathcal{P}_3^{(2,2)}(\bar{g}_3;\widehat{\bm{\sigma}})
	\equiv&
-\frac{1}{8\pi} \int_{-\infty}^\infty d\bar{g}_1 \int_{-\infty}^\infty d\bar{g}_2
	e^{-\bar{g}_1^2/2 - \bar{g}_2^2/2}
\bar{g}_2^2  U_{\beta 2} U_{\gamma 2} \Pi_{\beta\delta}\Pi_{\gamma\delta}\nonumber\\
=&-\frac{1}{4} U_{\beta 2} U_{\gamma 2}\Pi_{\beta\delta}\Pi_{\gamma\delta}\nonumber\\
=& -\frac{1}{4} \frac{1}{\widehat{\sigma}_x^2 + \widehat{\sigma}_y^2}
	\left\{
	\left(1-2\widehat{\sigma}_z^2\right) \Pi_{zz}^2
	+ \widehat{\sigma}_z^2\left[
	\widehat{\sigma}_x^2 \Pi_{xx}^2 + \widehat{\sigma}_y^2 \Pi_{yy}^2 + \widehat{\sigma}_z^2 \Pi_{zz}^2
	+ \left(\widehat{\sigma}_x^2+ \widehat{\sigma}_y^2\right)\Pi_{xy}^2 \right.\right.\nonumber\\
&\left.\left.+ 2\widehat{\sigma}_x \widehat{\sigma}_y \Pi_{xy}\left(\Pi_{xx} + \Pi_{yy}\right)
	\right]\right\},\\
\mathcal{P}_3^{(2,3)}(\bar{g}_3;\widehat{\bm{\sigma}})
	\equiv&
-\frac{1}{8\pi} \int_{-\infty}^\infty d\bar{g}_1 \int_{-\infty}^\infty d\bar{g}_2
	e^{-\bar{g}_1^2/2 - \bar{g}_2^2/2}
\bar{g}_3^2  U_{\beta 3} U_{\gamma 3} \Pi_{\beta\delta}\Pi_{\gamma\delta}\nonumber\\
=&-\frac{1}{4} \bar{g}_3^2 U_{\beta 3} U_{\gamma 3} \Pi_{\beta\delta}\Pi_{\gamma\delta}\nonumber\\
	=& -\frac{1}{4} \bar{g}_3^2
	\left[
	\widehat{\sigma}_x^2 \Pi_{xx}^2 + \widehat{\sigma}_y^2 \Pi_{yy}^2 + \widehat{\sigma}_z^2 \Pi_{zz}^2
	+ \left(\widehat{\sigma}_x^2+ \widehat{\sigma}_y^2\right)\Pi_{xy}^2 + 2\widehat{\sigma}_x \widehat{\sigma}_y \Pi_{xy}\left(\Pi_{xx} + \Pi_{yy}\right)
	\right].
\end{align}
\end{subequations}
Combining all the terms, we can obtain the expression of $\mathcal{P}_3^{(2)}(\bar{g}_3;\widehat{\bm{\sigma}})$ as
\begin{align}
	\mathcal{P}_3^{(2)}(\bar{g}_3;\widehat{\bm{\sigma}})
	=& -\frac{1}{4} \left(\bar{g}_3^2+\frac{\widehat{\sigma}_z^2}{\widehat{\sigma}_x^2 + \widehat{\sigma}_y^2}\right)
	\left[
	\widehat{\sigma}_x^2 \Pi_{xx}^2 + \widehat{\sigma}_y^2 \Pi_{yy}^2 + \widehat{\sigma}_z^2 \Pi_{zz}^2
	+ \left(\widehat{\sigma}_x^2+ \widehat{\sigma}_y^2\right)\Pi_{xy}^2 + 2\widehat{\sigma}_x \widehat{\sigma}_y \Pi_{xy}\left(\Pi_{xx} + \Pi_{yy}\right)
	\right]\nonumber\\
	&
	-\frac{1}{4} \frac{1}{\widehat{\sigma}_x^2 + \widehat{\sigma}_y^2}
	\left[
	\widehat{\sigma}_y^2 \Pi_{xx}^2 + \widehat{\sigma}_x^2 \Pi_{yy}^2 + \left(1-2\widehat{\sigma}_z^2\right) \Pi_{zz}^2
	+ \left(\widehat{\sigma}_x^2+ \widehat{\sigma}_y^2\right)\Pi_{xy}^2 - 2\widehat{\sigma}_x \widehat{\sigma}_y \Pi_{xy}\left(\Pi_{xx} + \Pi_{yy}\right)
	\right].
	\label{eq:P3_3}
\end{align}

Now we turn to the contribution $\mathcal{P}_3^{(3)}(\bar{g}_3;\widehat{\bm{\sigma}})$. It can be expressed as
\begin{equation}
	\mathcal{P}_3^{(3)}(\bar{g}_3;\widehat{\bm{\sigma}})
	= \sum_{i=1}^9 \mathcal{P}_3^{(3,i)}(\bar{g}_3;\widehat{\bm{\sigma}}),
\end{equation}
where
\begin{subequations}
\allowdisplaybreaks
\begin{align}
\mathcal{P}_3^{(3,1)}(\bar{g}_3;\widehat{\bm{\sigma}})
	\equiv& \frac{1}{32\pi} \int_{-\infty}^\infty d\bar{g}_1 \int_{-\infty}^\infty d\bar{g}_2
	e^{-\bar{g}_1^2/2 - \bar{g}_2^2/2}
\bar{g}_1^4 U_{\beta 1} U_{\gamma 1} U_{\delta 1} U_{\mu 1}\Pi_{\beta\gamma}\Pi_{\delta\mu}\nonumber\\
=&\frac{3}{16} U_{\beta 1} U_{\gamma 1} U_{\delta 1} U_{\mu 1}\Pi_{\beta\gamma}\Pi_{\delta\mu}\nonumber\\
=& \frac{3}{16}\frac{1}{\left(\widehat{\sigma}_x^2 + \widehat{\sigma}_y^2\right)^2}
		\left[\widehat{\sigma}_\beta \widehat{\sigma}_\gamma \Pi_{\beta\gamma}-\left(1-2\widehat{\sigma}_z^2\right)\left(\Pi_{xx}+\Pi_{yy}\right)\right]^2,\\
\mathcal{P}_3^{(3,2)}(\bar{g}_3;\widehat{\bm{\sigma}})
	\equiv& \frac{1}{32\pi} \int_{-\infty}^\infty d\bar{g}_1 \int_{-\infty}^\infty d\bar{g}_2
	e^{-\bar{g}_1^2/2 - \bar{g}_2^2/2}
\bar{g}_2^4 U_{\beta 2} U_{\gamma 2} U_{\delta 2} U_{\mu 2}\Pi_{\beta\gamma}\Pi_{\delta\mu}\nonumber\\
=& \frac{3}{16} U_{\beta 2} U_{\gamma 2} U_{\delta 2} U_{\mu 2}\Pi_{\beta\gamma}\Pi_{\delta\mu}\nonumber\\
=&\frac{3}{16}\frac{1}{\left(\widehat{\sigma}_x^2 + \widehat{\sigma}_y^2\right)^2}
		\left[\widehat{\sigma}_z^2\widehat{\sigma}_\beta \widehat{\sigma}_\gamma \Pi_{\beta\gamma}-\left(1-2\widehat{\sigma}_z^2\right)\left(\Pi_{xx}+\Pi_{yy}\right)\right]^2,\\
\mathcal{P}_3^{(3,3)}(\bar{g}_3;\widehat{\bm{\sigma}})
	\equiv& \frac{1}{32\pi} \int_{-\infty}^\infty d\bar{g}_1 \int_{-\infty}^\infty d\bar{g}_2
	e^{-\bar{g}_1^2/2 - \bar{g}_2^2/2}
\bar{g}_3^4 U_{\beta 3} U_{\gamma 3} U_{\delta 3} U_{\mu 3}\Pi_{\beta\gamma}\Pi_{\delta\mu}\nonumber\\
=&\frac{1}{16} \bar{g}_3^4 U_{\beta 3} U_{\gamma 3} U_{\delta 3} U_{\mu 3}\Pi_{\beta\gamma}\Pi_{\delta\mu}\nonumber\\
=&\frac{1}{16} \bar{g}_3^4 \left(\widehat{\sigma}_\beta \widehat{\sigma}_\gamma \Pi_{\beta\gamma}\right)^2,\\
\mathcal{P}_3^{(3,4)}(\bar{g}_3;\widehat{\bm{\sigma}})
	\equiv& \frac{1}{32\pi} \int_{-\infty}^\infty d\bar{g}_1 \int_{-\infty}^\infty d\bar{g}_2
	e^{-\bar{g}_1^2/2 - \bar{g}_2^2/2}
2 \bar{g}_1^2 \bar{g}_2^2 U_{\beta 1} U_{\gamma 1} U_{\delta 2} U_{\mu 2}\Pi_{\beta\gamma}\Pi_{\delta\mu}\nonumber\\
=&\frac{1}{8} U_{\beta 1} U_{\gamma 1} U_{\delta 2} U_{\mu 2}\Pi_{\beta\gamma}\Pi_{\delta\mu}\nonumber\\
=&-\frac{1}{8}\frac{1}{\left(\widehat{\sigma}_x^2 + \widehat{\sigma}_y^2\right)^2}
\left[\widehat{\sigma}_\beta \widehat{\sigma}_\gamma \Pi_{\beta\gamma}-\left(1-2\widehat{\sigma}_z^2\right)\left(\Pi_{xx}+\Pi_{yy}\right)\right]
\left[\widehat{\sigma}_z^2\widehat{\sigma}_\beta \widehat{\sigma}_\gamma \Pi_{\beta\gamma}-\left(1-2\widehat{\sigma}_z^2\right)\left(\Pi_{xx}+\Pi_{yy}\right)\right],\\
\mathcal{P}_3^{(3,5)}(\bar{g}_3;\widehat{\bm{\sigma}})
	\equiv& \frac{1}{32\pi} \int_{-\infty}^\infty d\bar{g}_1 \int_{-\infty}^\infty d\bar{g}_2
	e^{-\bar{g}_1^2/2 - \bar{g}_2^2/2}
4 \bar{g}_1^2 \bar{g}_2^2 U_{\beta 1} U_{\gamma 2} U_{\delta 1} U_{\mu 2}\Pi_{\beta\gamma}\Pi_{\delta\mu}\nonumber\\
=&\frac{1}{4} U_{\beta 1} U_{\gamma 2} U_{\delta 1} U_{\mu 2}\Pi_{\beta\gamma}\Pi_{\delta\mu}\nonumber\\
=&\frac{1}{4} \frac{\widehat{\sigma}_z^2}{\left(\widehat{\sigma}_x^2 + \widehat{\sigma}_y^2\right)^2}
	\left[\widehat{\sigma}_x \widehat{\sigma}_y \left(\Pi_{xx} -\Pi_{yy}\right) - \left(\widehat{\sigma}_x^2 - \widehat{\sigma}_y^2\right)\Pi_{xy}\right]^2,\\
\mathcal{P}_3^{(3,6)}(\bar{g}_3;\widehat{\bm{\sigma}})
	\equiv& \frac{1}{32\pi} \int_{-\infty}^\infty d\bar{g}_1 \int_{-\infty}^\infty d\bar{g}_2
	e^{-\bar{g}_1^2/2 - \bar{g}_2^2/2}
2 \bar{g}_1^2 \bar{g}_3^2 U_{\beta 1} U_{\gamma 1} U_{\delta 3} U_{\mu 3}\Pi_{\beta\gamma}\Pi_{\delta\mu}\nonumber\\
=&\frac{1}{8} \bar{g}_3^2 U_{\beta 1} U_{\gamma 1} U_{\delta 3} U_{\mu 3}\Pi_{\beta\gamma}\Pi_{\delta\mu}\nonumber\\
=&-\frac{1}{8} \frac{\bar{g}_3^2}{\widehat{\sigma}_x^2 + \widehat{\sigma}_y^2}
		\left[\left(\widehat{\sigma}_\beta \widehat{\sigma}_\gamma \Pi_{\beta\gamma}\right)^2-\left(1-2\widehat{\sigma}_z^2\right)\left(\Pi_{xx}+\Pi_{yy}\right)
		\widehat{\sigma}_\beta \widehat{\sigma}_\gamma \Pi_{\beta\gamma}\right],\\
\mathcal{P}_3^{(3,7)}(\bar{g}_3;\widehat{\bm{\sigma}})
	\equiv& \frac{1}{32\pi} \int_{-\infty}^\infty d\bar{g}_1 \int_{-\infty}^\infty d\bar{g}_2
	e^{-\bar{g}_1^2/2 - \bar{g}_2^2/2}
4 \bar{g}_1^2 \bar{g}_3^2 U_{\beta 1} U_{\gamma 3} U_{\delta 1} U_{\mu 3}\Pi_{\beta\gamma}\Pi_{\delta\mu}\nonumber\\
=&\frac{1}{4} \bar{g}_3^2 U_{\beta 1} U_{\gamma 3} U_{\delta 1} U_{\mu 3} \Pi_{\beta\gamma}\Pi_{\delta\mu}\nonumber\\
=&\frac{1}{4} \frac{\bar{g}_3^2}{\widehat{\sigma}_x^2+\widehat{\sigma}_y^2} \left[\widehat{\sigma}_x \widehat{\sigma}_y \left(\Pi_{xx} -\Pi_{yy}\right) - \left(\widehat{\sigma}_x^2 - \widehat{\sigma}_y^2\right)\Pi_{xy}\right]^2,\\
\mathcal{P}_3^{(3,8)}(\bar{g}_3;\widehat{\bm{\sigma}})
	\equiv& \frac{1}{32\pi} \int_{-\infty}^\infty d\bar{g}_1 \int_{-\infty}^\infty d\bar{g}_2
	e^{-\bar{g}_1^2/2 - \bar{g}_2^2/2}
2 \bar{g}_2^2 \bar{g}_3^2 U_{\beta 2} U_{\gamma 2} U_{\delta 3} U_{\mu 3}\Pi_{\beta\gamma}\Pi_{\delta\mu}\nonumber\\
=&\frac{1}{8} \bar{g}_3^2 U_{\beta 2} U_{\gamma 2} U_{\delta 3} U_{\mu 3} \Pi_{\beta\gamma}\Pi_{\delta\mu}\nonumber\\
=&\frac{1}{8} \frac{\bar{g}_3^2}{\widehat{\sigma}_x^2 + \widehat{\sigma}_y^2}
		\left[\widehat{\sigma}_z^2 \left(\widehat{\sigma}_\beta \widehat{\sigma}_\gamma \Pi_{\beta\gamma}\right)^2
		- \left(1-2\widehat{\sigma}_z^2\right) \left(\Pi_{xx}+\Pi_{yy}\right)
		\widehat{\sigma}_\beta \widehat{\sigma}_\gamma \Pi_{\beta \gamma}\right],\\
\mathcal{P}_3^{(3,9)}(\bar{g}_3;\widehat{\bm{\sigma}})
	\equiv& \frac{1}{32\pi} \int_{-\infty}^\infty d\bar{g}_1 \int_{-\infty}^\infty d\bar{g}_2
	e^{-\bar{g}_1^2/2 - \bar{g}_2^2/2}
4 \bar{g}_2^2 \bar{g}_3^2 U_{\beta 2} U_{\gamma 3} U_{\delta 2} U_{\mu 3}\Pi_{\beta\gamma}\Pi_{\delta\mu}\nonumber\\
=&\frac{1}{4} \bar{g}_3^2 U_{\beta 2} U_{\gamma 3} U_{\delta 2} U_{\mu 3} \Pi_{\beta\gamma}\Pi_{\delta\mu}\nonumber\\
=& \frac{1}{4} \frac{\bar{g}_3^2\widehat{\sigma}_z^2}{\widehat{\sigma}_x^2+\widehat{\sigma}_y^2}\left(\widehat{\sigma}_\beta \widehat{\sigma}_\gamma \Pi_{\beta\gamma}+\Pi_{xx}+\Pi_{yy}\right)^2.
\end{align}
\end{subequations}
After some algebra, one can obtain the combinations
\begin{subequations}
\begin{align}
&\mathcal{P}_3^{(3,1)}(\bar{g}_3;\widehat{\bm{\sigma}})
	+\mathcal{P}_3^{(3,2)}(\bar{g}_3;\widehat{\bm{\sigma}})
	+\mathcal{P}_3^{(3,4)}(\bar{g}_3;\widehat{\bm{\sigma}})
	+\mathcal{P}_3^{(3,5)}(\bar{g}_3;\widehat{\bm{\sigma}})\nonumber\\
	=& \frac{3}{16}\left(\widehat{\sigma}_\beta \widehat{\sigma}_\gamma \Pi_{\beta\gamma}\right)^2
	-\frac{1}{4} \frac{1}{\widehat{\sigma}_x^2 + \widehat{\sigma}_y^2}\left\{
	\left(\Pi_{xx}+\Pi_{yy}\right)\widehat{\sigma}_\beta \widehat{\sigma}_\gamma \Pi_{\beta\gamma}
	-\left(1-3\widehat{\sigma}_z^2\right) \left(\Pi_{xx}+\Pi_{yy}\right)^2\right.\nonumber\\
	&\left.-\widehat{\sigma}_z^2\left[\widehat{\sigma}_x^2 \Pi_{xx}^2 + \widehat{\sigma}_y^2 \Pi_{yy}^2 +\widehat{\sigma}_z^2 \Pi_{zz}^2
	+ \left(\widehat{\sigma}_x^2 +\widehat{\sigma}_y^2\right) \Pi_{xy}^2 + 2\widehat{\sigma}_x \widehat{\sigma}_y \Pi_{xy}\left(\Pi_{xx}+\Pi_{yy}\right)\right]\right\},\\
	&\mathcal{P}_3^{(3,6)}(\bar{g}_3;\widehat{\bm{\sigma}})
	+\mathcal{P}_3^{(3,7)}(\bar{g}_3;\widehat{\bm{\sigma}})
	+\mathcal{P}_3^{(3,8)}(\bar{g}_3;\widehat{\bm{\sigma}})
	+\mathcal{P}_3^{(3,9)}(\bar{g}_3;\widehat{\bm{\sigma}})\nonumber\\
	=& -\frac{3}{8} \bar{g}_3^2\left(\widehat{\sigma}_\beta \widehat{\sigma}_\gamma \Pi_{\beta\gamma}\right)^2
	+\frac{1}{4}\bar{g}_3^2\left[\widehat{\sigma}_x^2 \Pi_{xx}^2 + \widehat{\sigma}_y^2 \Pi_{yy}^2 +\widehat{\sigma}_z^2 \Pi_{zz}^2
	+ \left(\widehat{\sigma}_x^2 +\widehat{\sigma}_y^2\right) \Pi_{xy}^2 + 2\widehat{\sigma}_x \widehat{\sigma}_y \Pi_{xy}\left(\Pi_{xx}+\Pi_{yy}\right)\right].
\end{align}
\end{subequations}
Thus,
\begin{align}
	\mathcal{P}_3^{(3)}(\bar{g}_3;\widehat{\bm{\sigma}})
	&= \frac{\bar{g}_3^4-6\bar{g}_3^2+3}{16}\left(\widehat{\sigma}_\beta \widehat{\sigma}_\gamma \Pi_{\beta\gamma}\right)^2
	-\frac{1}{4} \frac{1}{\widehat{\sigma}_x^2 + \widehat{\sigma}_y^2}
	\left(\Pi_{xx}+\Pi_{yy}\right)\widehat{\sigma}_\beta \widehat{\sigma}_\gamma \Pi_{\beta\gamma}
	+\frac{1}{4} \frac{1-3\widehat{\sigma}_z^2}{\widehat{\sigma}_x^2 + \widehat{\sigma}_y^2} \left(\Pi_{xx}+\Pi_{yy}\right)^2\nonumber\\
	&\hspace{1em}+\frac{1}{4}\left(\bar{g}_3^2+\frac{\widehat{\sigma}_z^2}{\widehat{\sigma}_x^2 + \widehat{\sigma}_y^2}\right)
	\left[\widehat{\sigma}_x^2 \Pi_{xx}^2 + \widehat{\sigma}_y^2 \Pi_{yy}^2 +\widehat{\sigma}_z^2 \Pi_{zz}^2
	+ \left(\widehat{\sigma}_x^2 +\widehat{\sigma}_y^2\right) \Pi_{xy}^2 + 2\widehat{\sigma}_x \widehat{\sigma}_y \Pi_{xy}\left(\Pi_{xx}+\Pi_{yy}\right)\right].
	\label{eq:P3_4}
\end{align}

Summing up Eqs.\ \eqref{eq:P3_1}, \eqref{eq:P3_2}, \eqref{eq:P3_3}, and \eqref{eq:P3_4}, we finally obtain
\begin{equation}
	\mathcal{P}_3(\bar{g}_3;\widehat{\bm{\sigma}})
	= 1 + \frac{\bar{g}_3^2-1}{2} \widehat{\sigma}_\beta \widehat{\sigma}_\gamma \Pi_{\beta \gamma}
	+\frac{\bar{g}_3^4 - 6\bar{g}_3^2+3}{16} \left(\widehat{\sigma}_\beta \widehat{\sigma}_\gamma \Pi_{\beta \gamma}\right)^2.
	\label{eq:P3}
\end{equation}

\subsection{Evaluations of $I^{(\ell)}(\widehat{\bm{\sigma}})$ and $I_\alpha^{(\ell)}(\widehat{\bm{\sigma}})$}
Inserting Eqs.\ \eqref{eq:P31}, \eqref{eq:P32}, and \eqref{eq:P3} into  Eqs.\ \eqref{eq:def_I}, and performing the integrations over $\bar{g}_3$, we can obtain the expressions of $\tilde{I}^{(\ell)}$, $\tilde{\bar{I}}_1^{(\ell)}$, and $\tilde{\bar{I}}_2^{(\ell)}$.
Although we can derive their general expressions for arbitrary $\ell$, we are here interested in  $\tilde{I}^{(2)}$, $\tilde{I}^{(3)}$, $\tilde{\bar{I}}_1^{(2)}$, and $\tilde{\bar{I}}_2^{(3)}$.
These functions are given by Eqs.\ \eqref{eq:I2}, \eqref{eq:I3}, and
\begin{subequations}
\begin{align}
	\tilde{\bar{I}}_1^{(2)}(\widehat{\bm{\sigma}})
	=&  \frac{1}{\sqrt{\widehat{\sigma}_x^2+\widehat{\sigma}_y^2}}
	\left[\widehat{\sigma}_x \widehat{\sigma}_y \left(\Pi_{xx} -\Pi_{yy}\right) -\left(\widehat{\sigma}_x^2-\widehat{\sigma}_y^2\right)\Pi_{xy}\right]
	\left[\sqrt{\frac{2}{\pi}} e^{-b_T^2/2}\left(1+\frac{1}{4}\widehat{\sigma}_\beta \widehat{\sigma}_\gamma \Pi_{\beta\gamma}\right) - b_T {\rm erfc}\left(\frac{b_T}{\sqrt{2}}\right)
	 \right],\\
	\tilde{\bar{I}}_2^{(2)}(\widehat{\bm{\sigma}})
	= &\frac{\widehat{\sigma}_z}{\sqrt{\widehat{\sigma}_x^2+\widehat{\sigma}_y^2}}
	\left(\widehat{\sigma}_\beta \widehat{\sigma}_\gamma \Pi_{\beta\gamma} +\Pi_{xx} + \Pi_{yy}\right)
	\left[\sqrt{\frac{2}{\pi}} e^{-b_T^2/2}\left(1+\frac{1}{4}\widehat{\sigma}_\beta \widehat{\sigma}_\gamma \Pi_{\beta\gamma}\right) - b_T {\rm erfc}\left(\frac{b_T}{\sqrt{2}}\right)
	 \right].
\end{align}
\end{subequations}
Introducing  $\tilde{J}_\alpha \equiv U_{\alpha 1} \tilde{\bar{I}}_1^{(2)} + U_{\alpha 2} \tilde{\bar{I}}_2^{(2)}$, one obtains Eqs.\ \eqref{eq:Jx}--\eqref{eq:Jz}.
The quantities $\tilde{I}_\alpha^{(2)}$  are obtained by inserting Eqs.\ \eqref{I2-I3-Jx-Jy} into Eq.\ \eqref{C7}.



\begin{thebibliography}{99}
\bibitem{Barnes89}
	H. A. Barnes,
	J. Rheol. {\bf 33}, 329 (1989).
\bibitem{Mewis11}
	J. Mewis and N. J. Wagner,
	{\it Colloidal Suspension Rheology} (Cambridge University Press, New York, 2011).
\bibitem{Brown14}
	E. Brown and H. M. Jaeger,
	Rep. Prog. Phys. {\bf 77}, 040602 (2014).
\bibitem{Lootens05}
	D. Lootens, H. van Damme, Y. H\'{e}mar, and P. H\'{e}braud,
	Phys. Rev. Lett. {\bf 95}, 268302 (2005).
\bibitem{Cwalina14}
	C. D. Cwalina and N. J. Wagner,
	J. Rheol. {\bf 58}, 949 (2004).
\bibitem{Brilliantov04}
	B. V. Brilliantov and T. P\"{o}schel,
	\emph{Kinetic Theory of Granular Gases} (Oxford Univ. Press, Oxford, 2004).
\bibitem{Brey98} J. J. Brey, J. W. Dufty, C. S. Kim, and A. Santos,
	Phys. Rev. E {\bf 58}, 4638 (1998).
\bibitem{Garzo99}
	V. Garz\'{o} and J. W. Dufty,
	Phys. Rev. E {\bf 59}, 5895 (1999).
\bibitem{Lutsko05}
	J. F. Lutsko,
	Phys. Rev. E {\bf 72}, 021306 (2005).
\bibitem{Garzo13}
	V. Garz\'{o},
	Phys. Fluids, {\bf 25},  043301 (2013).
\bibitem{Garzo}
	V. Garz\'{o},
	\emph{Granular Gaseous Flows ---A Kinetic Theory Approach to Granular Gaseous Flows---} (Springer  Nature, Switzerland, 2019).
\bibitem{Tsao95}
	H.-W. Tsao and D. L. Koch,
	J. Fluid Mech. {\bf 296}, 211 (1995).
\bibitem{Sangani96}
	A. S. Sangani, G. Mo. H.-W. Tsao, and D. L. Koch,
	J. Fluid Mech. {\bf 313}, 309 (1996).
\bibitem{BGK2016}
	H. Hayakawa and S. Takada,
	EPJ Web Conf. {\bf 140}, 09003 (2017).
\bibitem{DST16}
	H. Hayakawa and S. Takada,
	Prog. Theor. Exp. Phys. {\bf 2019}, 083J01 (2019).
\bibitem{Saha17}
	S. Saha and M. Alam,
	J. Fluid. Mech. {\bf 833}, 206 (2017).
\bibitem{Chamorro15}
	M. G. Chamorro, F. Vega Reyes, and V. Garz\'{o},
	Phys. Rev. E {\bf 92}, 052205 (2015).	
\bibitem{Gonzalez20b}
	R. G\'{o}mez Gonz\'{a}lez, and V. Garz\'{o}, 
	Phys. Fluids \textbf{32}, 073315 (2020).
\bibitem{Hayakawa17}
	H. Hayakawa, S. Takada, and V. Garz\'{o},
	Phys. Rev. E {\bf 96}, 042903 (2017), [Erratum] Phys. Rev. E {\bf 101}, 069904 (2020).
\bibitem{Scala12}
	A. Scala,
	Phys. Rev. E {\bf 86}, 026709 (2012).
\bibitem{Koch01}
	D. L. Koch and R. J. Hill,
	Ann. Rev. Fluid Mech., {\bf 33}, 619 (2001).
\bibitem{Sugimoto20}
	S. Sugimoto and S. Takada,
	J. Phys. Soc. Jpn. {\bf 89}, 084803 (2020).
\bibitem{Grad49}
	H. Grad,
	Commun. Pure Appl. Math. {\bf 2}, 331 (1949).
\bibitem{Resibois77}
	P. R\'{e}sibois and M. de Leener,
	\emph{Classical Kinetic Theory of Fluids} (John Wiley $\&$ Sons, New York, 1978).
\bibitem{Chialvo13}
	S. Chialvo and S. Sundaresan,
	Phys. Fluids {\bf 25}, 070603 (2013).
\bibitem{K81}
	N. G. van Kampen, 
	\emph{Stochastic Processes in Physics and Chemistry} (North Holland, Amsterdam, 1981).
\bibitem{Kawasaki14}
	T. Kawasaki, A. Ikeda, and L. Berthier,
	EPL {\bf 107}, 28009 (2014).	
\bibitem{LE72}
	A. W. Lees and S. F. Edwards,
	J. Phys. C \textbf{5}, 1921 (1972).	
\bibitem{Hayakawa03}
	H. Hayakawa,
	Phys. Rev. E {\bf 68}, 031304 (2003).
\bibitem{CS}
	N. F. Carnahan and K. E. Starling,
	J. Chem. Phys. {\bf 45}, 2102 (1969).
\bibitem{Santos98}
	A. Santos, J. M. Montanero, J. W. Dufty, and J. J. Brey,
	Phys. Rev. E {\bf 57}, 1644 (1998).
\bibitem{Montanero99}
	J. M. Montanero, V. Garz\'{o}, A. Santos, and J. J. Brey,
	J. Fluid Mech. {\bf 389}, 391 (1999).
\bibitem{Gradenigo11}
	G. Gradenigo, A. Sarracino, D. Villamana, and A. Puglisi, 
	J. Stat. Mech. P08017 (2011).
\bibitem{Garzo12}
	V. Garz\'{o}, S. Tenneti, S. Subramaniam, and C. M. Hrenya, 
	J. Fluid Mech. {\bf 712}, 129 (2012).
\bibitem{Garzo13PF}
	V. Garz\'{o}, M. G. Chamorro, and F. Vega Reyes, 
	Phys. Rev. E {\bf 87}, 032201 (2013).
\bibitem{Gonzalez19}
	R. G\'{o}mez Gonz\'{a}lez, and V. Garz\'{o}, 
	J. Stat. Mech. 093204 (2019).
\bibitem{Khalil13}
	N. Khalil and V. Garz\'{o}, 
	Phys. Rev. E {\bf 88}, 052201 (2013).  
\bibitem{Gonzalez20}
	R. G\'{o}mez Gonz\'{a}lez, N. Khalil, and V. Garz\'{o}, 
	Phys. Rev. E {\bf 101}, 012904 (2020).	
\bibitem{Garzo02}
	V. Garz\'{o},
	Phys. Rev. E {\bf 66}, 021308 (2002).
\bibitem{Santos04}
	A. Santos, V. Garz\'{o}, and J. W. Dufty,
	Phys. Rev. E {\bf 69}, 061303 (2004).
\bibitem{LBD02}
	J. Lutsko, J. J. Brey, and J. W. Dufty,
	Phys. Rev. E {\bf 65}, 051304 (2002).
\bibitem{DHGD02}
	S. R. Dahl, C. M. Hrenya, V. Garz\'{o}, and J. W. Dufty,
	Phys. Rev. E {\bf 66}, 041301 (2002).
\bibitem{MGAL06}
	J. M. Montanero, V. Garz\'{o}, M. Alam, and S. Luding,
	Granul. Matter {\bf 8}, 103 (2006).
\bibitem{MDCPH11}
	P. Mitrano, S. R. Dhal, D. J. Cromer, M. S. Pacella, and C. M. Hrenya,
	Phys. Fluids {\bf 23}, 093303 (2011).
\bibitem{MGH14}
	P. P. Mitrano, V. Garz\'{o}, and C. M. Hrenya,
	Phys. Rev. E {\bf 89}, 020201(R) (2014).
\bibitem{Suzuki17}
	K. Suzuki and H. Hayakawa,
	J. Fluid Mech. {\bf 864}, 1125 (2019).
\bibitem{Suzuki15}
	K. Suzuki and H. Hayakawa,
	Phys. Rev. Lett. {\bf 115}, 098001 (2015).
\bibitem{Saitoh17}
	K. Saitoh and H. Hayakawa,
	in preparation.
\bibitem{Saitoh16}
	K. Saitoh and H. Hayakawa,
	EPJ Web Conf. {\bf 140}, 03063 (2017).
\bibitem{BHL84}
	F. G. Bridges, A. Hatzes, and D. N. C. Lin,
	Nature {\bf 309}, 333 (1984).
\bibitem{RPBS99}
	R. Ram\'{\i}rez, T. P\"{o}schel, N. V. Brilliantov, and T. Schwager,
	Phys. Rev. E {\bf 60}, 4465 (1999).
\bibitem{BP00}
	N. V. Brilliantov and T. P\"{o}schel,
	Phys. Rev. E {\bf 61}, 5573 (2000).
\bibitem{BP03}
	N. V. Brilliantov and T. P\"{o}schel,
	Phys. Rev. E {\bf 67}, 061304 (2003).
\bibitem{DBPB13}
	A. K. Dubey, A. Bodrova, S. Puri, and N. V. Brilliantov,
	Phys. Rev. E {\bf 87}, 062202 (2013).
\bibitem{Yamada75} 
	T. Yamada and K. Kawasaki, 
	Prog. Theor. Phys. {\bf 53}, 111 (1975).
\bibitem{Otsuki09a} 
	M. Otsuki and H. Hayakawa, 
	J. Stat. Mech. Theor. Exp. L08003 (2009).
\bibitem{Otsuki09b}
	M. Otsuki and H. Hayakawa, 
	Eur. Phys. J. Special Topics {\bf 179}, 179 (2009).	
\end{thebibliography}
\end{document}